\documentclass[prl,preprint,showpacs]{revtex4}
\usepackage{ae,aecompl}
\usepackage{helvet}
\usepackage{epsfig}
\usepackage{dcolumn}
\usepackage{color}
\usepackage[LGR,T1]{fontenc}
\usepackage[latin9]{inputenc}
\usepackage{amssymb}
\usepackage{amsmath}
\usepackage{esint}

\DeclareFontEncoding{LGR}{}{}
\DeclareTextSymbol{\~}{LGR}{126}

\begin{document}
\bibliographystyle{apsrev}

\title{Remarkable effects of disorder on superconductivity of single atomic layers of lead on silicon}

\author{C. Brun$^1$}
\author{T. Cren$^1$}
\email{tristan.cren@insp.jussieu.fr}
\author{V. Cherkez$^1$}
\author{F. Debontridder$^1$}
\author{S. Pons$^1$}
\author{L.B Ioffe$^2$}
\author{B.L. Altshuler$^3$}
\author{D. Fokin$^{1,4}$}
\author{M. C. Tringides$^5$}
\author{S. Bozhko$^6$}
\author{D. Roditchev$^{1,7}$}

\affiliation{$^1$Institut des Nanosciences de Paris, Universit\'{e} Pierre
et Marie Curie UPMC and CNRS-UMR 7588, 4 place Jussieu, 75252 Paris, France}
\affiliation{$^2$Laboratoire de Physique Th\'{e}orique et Hautes Energies, Universit\'{e} Pierre et Marie Curie UPMC and CNRS UMR 7589, 4 Place Jussieu, 75252 Paris, France}
\affiliation{$^3$Physics Department, Columbia University, New York, New York 10027, USA}
\affiliation{$^4$Joint Institute for High Temperatures, RAS 125412, Moscow, Russia}
\affiliation{$^5$Ames Laboratory-U.S. Department of Energy, and Department of Physics and Astronomy, Iowa State University, Ames, Iowa, 50011, USA}
\affiliation{$^6$Institute for Solid State Physics, RAS 142432, Chernogolovka, Russia}
\affiliation{$^7$LPEM-UMR8213/CNRS-ESPCI ParisTech-UPMC, 10 rue Vauquelin-75005 Paris, France}

\date{\today}

\begin{abstract}

\textbf{In bulk materials superconductivity is remarkably robust with respect to non-magnetic disorder \cite{Anderson}. In the two-dimensional limit however, the quantum condensate suffers from the effects produced by disorder and electron correlations which both tend to destroy superconductivity \cite{Goldman1998,Huscroft,Bouadim2011,Feigelman,Feigelman2,Sacepe2011}. The recent discovery of superconductivity in single atomic layers of Pb, the striped incommensurate (SIC) and $\sqrt{7}\times\sqrt{3}$ Pb/Si(111) \cite{Xue,Yamada}, opened an unique opportunity to probe the influence of well-identified structural disorder on two-dimensional superconductivity at the atomic and mesoscopic scale \cite{Uchihashi,FeSeXue}. In these two ultimate condensates we reveal \emph{how} the superconducting spectra loose their conventional character, by mapping the local tunneling density of states. We report variations of the spectral properties even at scales significantly shorter than the coherence length. Furthermore, fine structural differences between the two monolayers, such as their atomic density, lead to very different superconducting behaviour. The denser SIC remains globally robust to disorder, as are thicker Pb films \cite{Yang,Ozer,Eom,OzerAlloy,Brun,Qin}, whereas in the slighly more diluted $\sqrt{7}\times\sqrt{3}$ system superconductivity is strongly fragilized. A consequence of this weakness is revealed at monoatomic steps of $\sqrt{7}\times\sqrt{3}$, which disrupt superconductivity at the atomic scale. This effect witnesses that each individual step edge is a Josephson barrier. At a mesoscopic scale the weakly linked superconducting atomic terraces of $\sqrt{7}\times\sqrt{3}$ form a native network of Josephson junctions. We anticipate the Pb/Si(111) system to offer the unique opportunity to tune the superconducting coupling between adjacent terraces \cite{Devilstaircase}, paving a new way of designing atomic scale quantum devices compatible with silicon technology.}
\end{abstract}


\maketitle

The ideal striped incommensurate (SIC) monolayer (ML) contains 1.33 ML of Pb atoms on top of a Si(111) substrate \cite{Seehofer,Devilstaircase}, whereas the $\sqrt{7}\times\sqrt{3}$ Pb/Si(111) (shortened hereafter by $\sqrt{7}\times\sqrt{3}$) contains 1.20 ML \cite{Horikoshi,Kumpf}. Both systems have close atomic structure and differ by less than 10~\% in their Pb atom density \cite{Brochard,Cudazzo}. The two-dimensional (2D) metallic electronic state that carries surface superconductivity arises from the Pb-Si hybridization at the atomically sharp Pb/Si interface. This state rapidly decays inside Si-substrate; it is not coupled to electronic bands of bulk Si \cite{ARPES0,ARPES1,KangDFT}. The two studied superconductors have similar critical temperatures: 1.8K for the SIC and 1.5K for the $\sqrt{7}\times\sqrt{3}$ \cite{Xue}, suggesting a priori that these two systems possess close superconducting properties.

The SIC and $\sqrt{7}\times\sqrt{3}$ Pb/Si(111) monolayers were grown \emph{in situ} on regular atomic terraces of Si(111) reconstructed 7$\times$7 (see the Methods section and Supplementary Fig.S1 for the samples preparation and characterization). The atomically flat terraces of several hundreds of nanometers width are separated by monoatomic steps. Figs.1a,e show the topographies of the corresponding reconstructed lattices with atomic resolution. Both surface reconstructions take place in three different rotational domains oriented at 120$^{\circ}$ from each other. Different structural defects such as vacancies, adatoms, stacking faults inside a single domain and twin boundaries between different domains, are observed (Figs.1a,e).

Local tunneling conductance spectra acquired on atomic terraces at $T=320$mK (see the Methods section for details about the measurements) reveal in the SIC a fully developed gap centered at the Fermi level and pronounced coherence peaks at $V\approx\pm0.3$mV (Fig.1c). These are well-identified signatures of the elementary excitation spectrum of a conventional superconductor, as described by Bardeen-Cooper and Schrieffer (BCS) \cite{DeGennes}. The gap energy was extracted from BCS fits, without any additional broadening parameter, leading to $\Delta_{SIC}=0.23$meV. Despite a rather high density of atomic structural defects on terraces (Fig.1a), the superconducting gap in SIC is spatially homogeneous, as demonstrated by the corresponding zero-bias tunneling conductance map in Fig.1b; no specific bound states are observed at individual defects. Furthermore, it is remarkable to see how the zero-bias conductance remains insensitive to even larger structural defects, such as the two 1ML high nano-islands appearing very bright in the lower part of the topographic image presented in Figure 1a. These defects are completely invisible in the spectroscopic map Fig.1b, i.e. they do not create any subgap electronic states in accordance with the famous Anderson theorem \cite{Anderson}.

In the BCS framework, the Anderson theorem predicts no effect of the potential disorder on the coherence peak. In a disordered superconductor each one-particle excitation (quasiparticle) occupies a particular one-particle eigenstate (orbital) $a$ formed in a given realization of the disorder and characterized by an energy $\epsilon_{a}$ in the absence of the pairing interaction. Within the BCS approximation, occupation of each single-occupied orbital commutes with the pairing interaction and thus experiences no quantum fluctuations. Such orbitals are therefore \emph{blocked}, i.e. they do not participate in the formation of the superconducting gap, which is due to the interaction-induced mixing of double-occupied and empty states of the \emph{unblocked} orbitals (Fig.2a). A quasiparticle blocks one of the orbitals (e.g. $a$) and thus costs some interaction energy in addition to $\epsilon_{a}$, so its energy becomes $E_{a}=\sqrt{\epsilon_{a}^{2}+\Delta^{2}}$. This leads to the coherence peak singularity of the density of the states (DOS), $\nu(E)$ of the one-particle excitations at $E \approx \Delta$: $\nu(E)=\nu_{0}\sqrt{\Delta/2(E-\Delta)}$, where $\nu_{0}$ is the quasiparticle DOS at $E_F$. Thus the disorder can neither broaden nor reduce the quasiparticle peak as long as lons as $\nu_{0}$ remains non-zero.

We found that the tunneling spectroscopy of our extreme 2D SIC system contradicts this theoretical prediction. The spatial fluctuations of the amplitude of the coherence peak ($\sim20\%$ of relative change) manifest themselves on the spectroscopic map (Fig.1d) by the yellow-red (high amplitude) and blue-green (low amplitude) patches, not spatially correlated to the individual defects observed in Fig.1a. The sizes of the patches ($2-10$ nm) are much smaller than the coherence length $\xi_{SIC}\approx50$ nm \cite{Xue}. The red and blue curves on Fig.1c are representative tunneling spectra of each type of patches. The two curves show no subgap states and are characterized by the same gap $\Delta$. They are practically indistinguishable at low energies. However the heights of the coherence peaks in the different patches differ substantially. While the spectra acquired on yellow-red patches are in satisfactory agreement with the BCS predictions, the observed reduction of peak amplitude with a vanishing subgap DOS cannot be accounted for by BCS theory.

The common wisdom attributes the deviations from pure BCS behavior to pair-breaking, which can be due to the violation of the time-reversal symmetry or to some inelastic processes, e.g. electron-electron collisions \cite{Altshuler1982,Altshuler1985}. In terms of the one-particle orbitals the pair-breaking means blocking two orbitals by transferring one electron between them. Strong enough pair-breaking ($\Gamma>\Delta$) prevents the formation of the superconducting state. For weak pair-breaking ($\Gamma\ll\Delta$) superconductivity survives but the coherence peak in the tunneling DOS broadens and the subgap DOS becomes finite. The stronger the disorder the more pronounced both effects become.

The SIC spectra require a new interpretation. In this material the non-BCS part of the electron-electron interaction causes virtual processes that add to the real  pair-breaking. These processes mix blocked and unblocked states of the orbitals and modify the many body eigenstates of the system. This modification is usually neglected. However in other fermionic systems tending to condense (e.g. quantum atomic gases) this is the interaction that eventually drives the system to the Bose-condensed (BEC) state of two-fermion molecules. Indeed, two fermions in their bound state with high probability occupy different orbitals. Therefore the studies of these effects should shed some light on the BCS-BEC crossover, which is far from being well understood.

There are two reasons why the ``virtual pair-breaking\textquotedblright{} are usually neglected in weakly disordered superconductors. The first one is the smallness of the non-BCS matrix elements \cite{Aleiner2000}: in 2D systems the mixing between the blocked and unblocked orbitals is smaller than the mixing between the unblocked orbitals by the parameter $G^{-1}$, where $G$ is the dimensionless Thouless conductance \cite{Thouless1974}, $G=2\pi\hbar/(e^{2}R_{\square})$. The second reason is the large energy cost of each Cooper pair breaking, which further reduces the energy corrections. This is not the case for the non-BCS correction to the quasiparticle energy, which turns out to be $\Delta/E\gg1$ times larger than the correction to the ground state. Indeed, the electron which tunnels into a superconductor with all orbitals unblocked, blocks one of the orbitals ($\alpha$) and within the BCS approximation resides there forever. The resulting state $\left|\alpha\right\rangle $ is an eigenstate of the BCS Hamiltonian. The non-BCS interaction will cause the virtual transitions of the electron between the orbitals and create a linear combination of these states $\left|a\right\rangle +\sum_{b\neq a}c_{b}\left|b\right\rangle $ with the modified energy $\tilde{E}_{a}=\sqrt{\Delta^{2}+\epsilon_{a}^{2}}+\delta E_{a}$. In the lowest order of the perturbation theory in the non-BCS interaction ($\propto1/G^{2}$) $\delta E_{a}<0$ for the ground state ($\epsilon_{a}=0$) and increases with $\epsilon_{a}$. This means that the coherence peak in the tunneling DOS broadens and reduces its amplitude while subgap states do not appear.  The semi-quantitative analysis presented in the supporting materials suggests that the coherence peak is indeed broadened on the energy scale $\epsilon_{\star}=(C/G)^{4/3}\Delta$ where $C\approx7$ and the tunneling DOS has the form
\[
\nu(E)=\nu_{0}\left(\frac{G}{C}\right)^{2/3}F\left(\frac{E-\Delta}{\epsilon_{\star}}\right)
\]

The function $F(x)$ tends to $x^{-1/2}$ at $x\gg1$ which results in a conventional square root behavior at $E-\Delta\gg\epsilon_{\star}$. The singularity saturates at $x\sim1$ where $F(x)$ has a broad maximum resulting in $\nu_{max}\approx\nu_{0}(G/C)^{2/3}$. At lower energies the tunneling DOS decreases sharply and vanishes at $E=\Delta-2.02\epsilon_{\star}$. For lower energies the DOS is \emph{exactly} zero. Our model provides a good fit of the experimental data (Fig.2b, see supplementary material for more information on the theory).

In the $\sqrt{3}\times\sqrt{7}$ system the departure from the pure BCS behavior is yet stronger than in SIC. As Fig.1g demonstrates the coherence peaks are everywhere much lower and broader than the pure BCS theory would predict. The gap edge is less sharp and subgap filling is remarkable. Both the zero-bias conductance (Fig.1f) and coherence peak amplitude (Fig.1h) vary from point to point on a scale, which is smaller than the coherence length $\xi_{\sqrt{3}\times\sqrt{7}}\approx45$ nm. These spatial variations are not correlated with twin boundaries or local atomic defects observed in the topography (Fig.1e). Even in the most favorable regions where the coherence peak is the highest, the tunneling spectrum (red curve on Fig.1g) differs substantially from the pure BCS prediction (dotted line on Fig. 1g). However, adding to the BCS theory the pair-breaking with $\Gamma=0.02\, meV\approx0.1\Delta_{\sqrt{3}\times\sqrt{7}}$ (solid line on Fig.1g) provides an acceptable fitting of the experimental curve. The observed spectral variations on a length scale comparable to $\xi_{\sqrt{3}\times\sqrt{7}}$ should be caused mainly by inelastic processes, as significant magnetic disorder is excluded here. Furthermore, a large Rashba term is expected in our atomically thin heavy materials \cite{DilRashbaQWS,SetphanePons,RashbaMLPb,DopRashbaQWS}. This should lead to a significant momentum dependence of the gap \cite{Gor'kovRashba} that would result in an effective magnetic behaviour of the conventional non-magnetic impurities. Thus, we anticipate the observed short-scale spectral variations on lengths shorter than $\xi_{\sqrt{3}\times\sqrt{7}}$ as being due to the scattering of the triplet part of the order parameter by the non magnetic disorder of the system. This Rashba scattering component contributes also to the total effective $\Gamma=0.02$ meV.

Further differences between the superconducting properties of our two monolayer-superconductors are evidenced in a spectacular manner at monoatomic step edges, where two neighboring terraces join. Two typical regions of each sample, centered at a monoatomic step edge, are shown in Fig.3a,c. In the vicinity of the structural perturbation represented by the step edge, the superconductivity in the SIC remains unperturbed, as the zero-bias conductance map emphasizes (Fig.3b); across the step edge, the tunneling spectra remain identical to those measured on terraces, rendering the step edge completely invisible in the zero-bias conductance map. In contrast, abrupt spectral changes are observed when crossing a step of the $\sqrt{7}\times\sqrt{3}$. They appear in a narrow energy window ($\pm0.5$mV around zero-bias) and are thus related to the superconducting gap. In Fig.3d, these changes correspond to abrupt jumps of the zero-bias conductance from the lower terrace (blue color) to the upper one (yellow color). As Fig.3e emphasizes on a representative line scan, a discontinuity occurs right at the step edge, indicated by the vertical red arrow, between the spectra measured on the lower terrace and on the upper one. This discontinuity takes place on a length scale of few angstroms. Strikingly, the spectra measured on the lower terrace very close to the step edge are similar to the spectra taken farther away on the same lower terrace. On the contrary, Figs.3d,e show that most of the regions located on the upper terrace very close to the step edge (regions appearing in yellow in Fig.3d) present no fully developed superconducting gap but only a dip at the Fermi level. The revealed effect points to a sudden disruption of the superconducting order at a scale much shorter than $\xi_{\sqrt{7}\times\sqrt{3}}$. This kind of discontinuity occurs only in junctions formed of two superconductors or of a superconductor and a normal metal, separated by a barrier \cite{DeGennes}. Depending on the transparency of the interface barrier, there is a crossover from strong coupling regime to weak links as in Josephson junctions \cite{Josephson}. The observed abrupt spectral modifications demonstrate that individual step edges of the $\sqrt{7}\times\sqrt{3}$ system are strong barriers for Cooper pairs. This result furnishes a microscopic proof of similar conclusions drawn from macroscopic transport measurements on the $\sqrt{7}\times\sqrt{3}$ In/Si(111) system \cite{Uchihashi}. It means that the superconducting wave function is confined at the very surface since it cannot bypass through the bulk the surface barrier represented by the step edge. This observation demonstrates the pure surface nature of superconductivity in $\sqrt{7}\times\sqrt{3}$ Pb/Si(111).  


We now discuss how one can rationalize that in the $\sqrt{7}\times\sqrt{3}$ sample most regions situated on the upper terrace next to the step edge reveal dip-like spectra. The experimental data show that such regions are not intrinsically superconducting and are subjected to proximity effect from the neighboring upper terrace \cite{McMillan,DeGennes}. Moreover, a careful analysis shows that such regions are nano-protrusions which structure deviates from the perfect  $\sqrt{7}\times\sqrt{3}$ (see the Methods section, the parts 1, 3 and the Fig.S1 and S2 of the supplementary material, for more detailed information about nano-protrusions). As an exemple, their delimitation is shown in the topography of Fig.3c. Following the trajectory indicated by the dashed line in Fig.3d, when approaching the nano-protrusion from the upper terrace side, it is seen that the superconducting gap is progressively filled with quasiparticle states. Fig.3e allows a better appreciation of the smooth evolution of the zero-bias conductance and of the conductance spectra along this line. This phenomenon takes place on a scale of few tens of nanometers and leads to the absence of a well-developed gap close to the step edge, giving a spectroscopic evidence of the non-superconducting character of the nano-protrusions. Thus, Cooper pairs diffuse from upper superconducting terraces to these normal regions where they induce superconducting correlations \cite{SerrierGarcia,Kim}. Reciprocally, the superconductivity in terraces is weakened in the vicinity of these non-superconducting areas. This inverse proximity effect extends on the scale of the superconducting coherence length $\xi_{\sqrt{7}\times\sqrt{3}}$. From the spatial dependence of the inverse proximity effect, we estimate $\xi_{\sqrt{7}\times\sqrt{3}} \simeq 40$ nm. Additionally, the smooth spatial evolution of the spectra witnesses for a high electronic transparency between nano-protrusions and upper terraces. The structure and electronic properties of nano-protrusions are further discussed in the supplementary material.

The different role played by disorder in the SIC and $\sqrt{7}\times\sqrt{3}$ superconductors was further confirmed by the study of superconducting vortices. Vortices are present in most of superconductors subject to magnetic field as a direct consequence of the long range phase coherence of the quantum condensate \cite{Abrikosov}. The order parameter vanishes in the vortex centers, enabling scanning tunneling spectroscopy to reveal vortex cores as regions in which the gap is suppressed \cite{Hess}. In the SIC samples the vortices were easily observed (see Fig.4 and the part 3 of the supplementary material for more information). The cores present a conventional Abrikosov round shape \cite{Xue}. The half width at half maximum of the zero-bias conductance profile across single vortices leads to a coherence length $\xi_{SIC} \simeq 50$~nm, in agreement with \cite{Xue} but smaller than extrapolated from transport measurements \cite{Yamada}. Vortices are located both on terraces or near step edges without a clear preference. The vortex lattice is not perfectly triangular though it is characterized by a regular inter-vortex spacing.

The magnetic field response of the $\sqrt{7}\times\sqrt{3}$ system was found to be very different owing to its fragilization. In figure 5 we present three conductance maps acquired in the same sample area at different magnetic fields. When the field is increased, superconductivity is progressively suppressed, without clear signatures of vortices. The suppression first takes place at step edges (Fig.5b), at locations where already at zero-field the nano-protrusions weaken superconductivity (Fig.5a). These observations allow us to confirm previous conclusions : the step edges are Josephson weak links between adjacent atomic terraces. At higher fields (Fig.5c), the order parameter is further reduced in the step edges regions and simultaneously, several distorted vortices can be identified both in the interior of the main terrace region and on the decorated stacking fault lines of the underlying Si(111) substrate.

Due to the presence of non-superconducting regions already at zero-field, the elucidation of the vortex phase is a complex task. A deeper analysis of the tunneling data was done. Figures 5d-e show two maps, each of them being a difference between two successive maps of the upper panel. These difference maps enable a precise understanding of the spatial evolution of superconductivity with magnetic field. The first difference map (Fig.5d) shows that between $B_1=0$mT and $B_2=40$mT, the field-induced weakening of superconductivity along the steps and substrate stacking fault lines, is not homogeneous. Rather, in 7 specific regions (appearing in red) the gap is completely suppressed. These regions are all located at weak links and extend over a typical size of $\xi_{\sqrt{7}\times\sqrt{3}}$. Taking into account the fact that between $B_1$ and $B_2$ the magnetic flux over the scanned area has increased by $\delta\Phi_{1\rightarrow2}\approx7\Phi_0$, one can identify these 7 regions as mixed Abrikosov-Josephson vortices \cite{Gurevich}. The second difference map (Fig.5e) shows the evolution at higher fields, between $B_2=40$mT and $B_3=80$mT. It exhibits 6 more vortices, close to the expected flux variation $\delta\Phi_{2\rightarrow3}\approx7\Phi_0$. These additional vortices are now all located in the $\sqrt{7}\times\sqrt{3}$ terraces.

We show now that in $\sqrt{7}\times\sqrt{3}$ the possibility for the vortices to be located inside terraces strikingly depends on the terrace width. Fig.6 presents another sample region characterized by smaller terrace widths. In this area we find that for $B=40$mT the suppression occurs along step edges and decorated substrate stacking fault lines (see Fig.6c), as already reported in Fig.6b. However at higher field (Fig.6d), in strong contrast with Fig.5c, the order parameter is further reduced \emph{only} in the step edges regions where nano-protrusions are absent (areas indicated by arrows in Fig.6), which at zero-field showed no sign of discontinuity in the tunneling spectra (Fig.6b), and along the stacking fault lines of the Si(111) substrate. These observations allow us to extend our previous conclusions : both the step edges and the decorated stacking fault lines of the substrate, are Josephson weak links between adjacent atomic terraces. Finally, at $B_C=0.20-0.24$T superconductivity is suppressed everywhere (Fig.6f), enabling an independent consistent estimate of $\xi_{\sqrt{7}\times\sqrt{3}}=44$nm  ($B_C\approx\Phi_0/(2\pi\xi^2)$, $\Phi_0=h/{2e}$ being the flux quantum).

To summarize, we have studied by scanning tunneling spectroscopy the role played by disorder on two extreme superconductors consisting of one atomic layer of Pb grown on Si(111). The systems under study, referred to as the SIC and the $\sqrt{7}\times\sqrt{3}$, differ from each other only by 10~\% in their surface Pb atom density. We have measured and analyzed the spatial evolution of the local conductance spectra in the vicinity of various structural defects, and could link these observations to the superconducting behaviour at atomic and mesoscopic scale. For both systems, superconductivity in atomic terraces is found to be insensitive to individual non magnetic defects or twin boundaries. Nevertheless, global structural disorder affects both condensates. For the more robust SIC superconductor, this leads to spatial fluctuations of the coherence peaks height on a length scale much shorter than the coherence length; this new phenomenon can be explained theoretically by taking into account non-BCS corrections corresponding to virtual pair-breaking processes. In $\sqrt{7}\times\sqrt{3}$, the effects produced by disorder are strongly enhanced. This leads to large local smearing of the coherence peaks height, associated with important subgap filling. We attribute the spectral variations on a length scale comparable to the coherence length to pair-breaking effect induced by electron-electron interaction. As a consequence of a Rashba term, the spectral fluctuations occuring at shorter length scale are attributed to the scattering of the triplet part of the order parameter by non magnetic disorder. The weakness of the $\sqrt{7}\times\sqrt{3}$ superconductor is further revealed in the extreme sensitivity of this quantum condensate to monoatomic steps at the surface. Such structural defects provoke disruptions of the superconducting order at the atomic scale. In magnetic field, superconductivity preferentially weakens at these specific defects, which also pin vortices. Consequently, the whole $\sqrt{7}\times\sqrt{3}$ system is identified as a network of purely two-dimensional superconducting mesoscopic domains, interconnected at step edges and stacking faults lines of the substrate by atomically short Josephson links. The observed phenomenon is in striking contrast with a more conventional superconductivity in the SIC, which remains quite insensitive to steps and stacking faults lines of the substrate.

We emphasize that Pb/Si(111) is known to show a continuous structural crossover from the $\sqrt{7}\times\sqrt{3}$ Pb/Si(111) to the SIC phase \cite{Devilstaircase}, which is accessed by simply tuning the Pb atoms content. Thus, the Pb/Si(111) system offers the unique opportunity to control the superconducting links between adjacent terraces continuously, from weak to strong Josephson coupling. Our finding paves a new way of designing novel atomic-scale quantum devices involving pre-patterned or vicinal substrates, compatible with silicon technology.


\vspace{\baselineskip}
\textbf{Methods}

The 7$\times$7 reconstructed \textit{n}-Si(111) ($n\approx10^{19}cm^{-3}$) was prepared by direct current heating to 1200$^\circ$C followed by annealing procedure between 900$^\circ$C and 500$^\circ$C. Subsequently, 1.65 monolayer of Pb was evaporated on the Si(111)-7$\times$7 kept at room temperature, using an electron beam evaporator calibrated with a quartz micro-balance. The $\sqrt{7}\times\sqrt{3}$ Pb overlayer was formed by annealing at 230$^\circ$C for 30 minutes. The striped incommensurate phase (SIC) was grown by adding about 0.2 monolayer of Pb onto the $\sqrt{7}\times\sqrt{3}$ Pb/Si(111) reconstructed surface held at room temperature. At any stage of the sample preparation the pressure did not exceed $P = 3\times10^{-10}$ mbar. The sample structure was controlled in both real and reciprocal space by scanning tunneling microscopy and Low Energy Electron Diffraction. The scanning tunneling spectroscopy measurements were performed \textit{in situ} with a homemade apparatus, at a base temperature of 320mK and in ultrahigh vacuum $P < 3\times10^{-11}$ mbar \cite{Cren}; the electronic temperature was estimated to be 390mK. Mechanically sharpened PtIr tips were used. The bias voltage was applied to the sample with respect to the tip.  Typical set-point parameters for spectroscopy are 150 pA at $V=-5$mV. The tunneling conductance curves $dI(V)/dV$ were numerically differentiated from raw $I(V)$ experimental data. Each conductance map is extracted from a set of data consisting of spectroscopic $I(V)$ curves (256 data points per spectrum) measured at each point of a 256x256 grid, acquired simultaneously with the higher resolution topographic image. The magnetic field was applied perpendicular to the sample surface in the field-cooled regime.

\vspace{\baselineskip}
\textbf{Supplementary Information}

1. \textit{Sample characterization}. Pb monolayers were grown \emph{in situ} on regular atomic 7$\times$7 reconstructed terraces of Si(111) (see Fig.S1a) (see also the Methods section). The substrate exhibited regularly spaced 150-300nm wide atomically clean terraces separated by single atomic steps. Well ordered 7$\times$7 Si(111) reconstruction was routinely observed by scanning tunneling microscopy and confirmed by Low Energy Electron Diffraction patterns acquired at room temperature. On each Si(111) terrace atomically narrow stacking fault lines were revealed (pointed by arrows in Fig.S1a). They correspond to natural boundaries between neighboring domains of Si(111)-7$\times$7 shifted by 1/3 of the unit cell. The $\sqrt{7}\times\sqrt{3}$ Pb/Si(111) contains 1.20ML of Pb atoms, one monolayer being defined as the surface atomic density of Si(111) \cite{Horikoshi,Kumpf}. Upon Pb reconstruction, the atomic cleanness of the terraces is preserved in both Pb/Si systems, and the average width of terraces remains unchanged (see Fig.S1b). On each individual terrace, the chain-like $\sqrt{7}\times\sqrt{3}$ atomic structure was observed (Fig.S1-c1)); due to the 3-fold symmetry of the substrate, it is organized in three different rotational domains oriented in plane at 120$^{\circ}$ and corresponding to different crystallographic orientations of the $\sqrt{7}\times\sqrt{3}$, separated by twin boundaries (Fig.S1-c2) \cite{Horikoshi,Kumpf,Xue}. Other local structural defects such as voids, atom vacancies, adatoms, clusters/nanoislands and stacking fault lines were observed on terraces and identified (Fig.S1).

After Pb deposition and reconstruction the step edges in both Pb/Si samples appear more irregular than those of the initial 7$\times$7-Si(111) substrate. A closer examination of these irregularities allowed us to identify small nano-protrusions which grew at step edges and thus are attached to the upper terraces (see Fig.S1-c3). The supplementary figure S2 allows us to better emphasize how we identify precisely the location of the overhanging nano-protrusions, by following the necklace of small Pb clusters/nano-islands lying along the step edges. This explains how we could easily draw the boundary of the nano-protrusions in the Fig.3c of the main text. Other nano-protrusions decorate the staking fault lines of the Si(111)-7$\times$7 substrate and appear as one monolayer thick nano-islands (Fig.S1-b,c4). The nano-protrusions are reconstructed in both systems yet in $\sqrt{7}\times\sqrt{3}$ they show a significantly higher density of atomic defects and twin boundaries than the terraces. The lateral size of the nano-protrusions in SIC is smaller than in $\sqrt{7}\times\sqrt{3}$. In addition, as the SIC was grown by adding 0.20ML of Pb to the $\sqrt{7}\times\sqrt{3}$, this resulted in a slight extra amount of lead atoms ($\approx$0.07ML) with respect to the nominal reported coverage of the SIC (1.33ML). This led to the formation of very small 1ML high nano-islands on terraces (see Fig.4a of the main text). In both SIC and $\sqrt{7}\times\sqrt{3}$, nano-protrusions are attached to step edges and nano-islands grown on terraces were observed in previous reports \cite{Xue} and are thus characteristic structural features of these monolayer systems.

The structure of the SIC and $\sqrt{7}\times\sqrt{3}$ phases is different. The density of the $\sqrt{7}\times\sqrt{3}$ phase is 1.2ML. It is the same as the density of metallic Pb(111). This phase is thermally less stable than the SIC phase because it transforms to the 1x1 phase at 270 K. Its atomic structure is shown in the left panel of supplementary Fig.S3b. The $\sqrt{7}\times\sqrt{3}$ is a linear phase with rows along the $\bar{1}\bar{1}2$ direction which is the direction normal to the steps. This phase has 3-fold symmetry. Thus the Pb rows are oriented either perpendicular to the step edge, or form a 60 degree angle with respect to the step normal, as shown in supplementary Fig.S3a. Inside a single domain, the anisotropic structural properties of the $\sqrt{7}\times\sqrt{3}$ phase (linear phase) are reflected in its electronic properties (as measured by ARPES and calculated by DFT, see ref.\cite{Brochard,Cudazzo,ARPES0,ARPES1,KangDFT}).

The SIC phase has density approximately 1.28-1.3ML, less than 1.33ML the ideal coverage of the $\sqrt{3}\times\sqrt{3}$ phase and it develops out of the HIC phase of lower density 1.24-1.27ML. Structurally, the observed SIC phase is a mixed phase. It is made out of Pb triangular domains of the $\sqrt{3}\times\sqrt{3}$ phase (3-fold symmetry) having density 4/3ML, and meandering "filaments" of the $\sqrt{7}\times\sqrt{3}$ phase with density 1.2ML. The atomic structures of both phases are shown in the left and right panels of supplementary Fig.S3b. The SIC phase is stable up to 360K. Fig.S3b shows the unit cells of the two phases, small circles correspond to Si and bigger circles to Pb atoms; the different color corresponds to the Pb binding site on the Si substrate. Regarding the electronic properties of the SIC phase, they are expected to be more isotropic than those of the $\sqrt{7}\times\sqrt{3}$ phase (Ref.\cite{Devilstaircase,Hupalo2002}). Combined with the fact that the SIC is denser, its more isotropic electronic properties should help to understand why it is easier for the SIC phase than the $\sqrt{7}\times\sqrt{3}$ phase to cross the substrate steps. More detailed electronic calculations are needed to address quantitatively this issue.

For both phases we give in the table~\ref{defects} the density of each type of defects. The main difference is the density of domain boundaries. In other terms, the domains are in average larger in the $\sqrt{7}\times\sqrt{3}$ phase than in the SIC phase.

\begin{table}
\caption{\label{defects} Averaged density of defects as measured by STM for the SIC and $\sqrt{7}\times\sqrt{3}$ phases.}
\begin{ruledtabular}
\begin{tabular}{l|c|c|c|c|c|c|c}
Defects  &step       &atom       &other     &voids      &clusters/ &stacking &domain\\
density  &edges      &vacancies  &point     &$(nm^{-2})$&islands   &faults   &boundaries\\
         &$(nm^{-1})$&$(nm^{-2})$&defects   &           &$(nm^{-2})$&$(nm^{-1})$&$(nm^{-2})$\\
         &           &           &$(nm^{-2})$&          & &\\ \hline
$\sqrt{7}\times\sqrt{3}$&$5$ $10^{-3}$&$1-4$ $10^{-2}$&$3-6$ $10^{-2}$&$1-4$ $10^{-3}$&$3-7$ $10^{-4}$&$2$ $10^{-1}$&$1-3$ $10^{-2}$ \\
SIC&$7$ $10^{-3}$&$1-4$ $10^{-2}$&$1-4$ $10^{-1}$&$4-6$ $10^{-3}$&$4-7$ $10^{-3}$&$1.5-2.5$ $10^{-1}$&$0.6-2$ $10^{-1}$
\end{tabular}
\end{ruledtabular}
\end{table}

\vspace{\baselineskip}
2. \textit{Data processing}. The $dI/dV$ spectra are independent of the tunneling set point which was chosen far from the superconducting gap region (about ten to twenty times the bias voltage corresponding to the superconducting gap). The tunneling range used to measure the superconducting gap was 25-250 M$\Omega$. For the $\sqrt{7}\times\sqrt{3}$ phase, the fits are performed manually using a standard BCS DOS including an extra empirical pair-breaking parameter, the gamma parameter in the Dynes formula. Our purpose is not to provide a quantitative analysis but rather to emphasize the large deviations from BCS. The numerical differentiation is a standard centered three-points derivation. Then a smoothing algorithm was used to reduce the noise signal without affecting the spectral features inside the gap region. For this purpose we use a Gaussian variable filter. This procedure was possible because our $dI/dV$ spectra are measured with enough points in the spectra (less than 5.7 microVolts per point $\approx kT/5 \approx 2\Delta / 90$) to allow further proper filtering without loosing information.

\vspace{\baselineskip}

3. \textit{Scanning tunneling spectroscopy of nano-protrusions}. Scanning tunneling spectroscopy revealed different electronic characteristics of the nano-protrusions attached to the edges of upper terraces for the two systems. In the SIC sample, the local tunneling spectra exhibit a well developed superconducting gap everywhere across the step edge. Both the gap width and the overall shape of the spectra are identical to those acquired directly on the SIC terraces. Consequently, the nano-protrusions are indistinguishable in spectroscopic maps and thus are identified as superconducting (the reason for this could be either an intrinsic superconductivity of the nano-protrusion or a very strong induced proximity effect). As a result, both nano-protrusions at step edges and nano-islands located on terraces do not perturb the superconducting order in the SIC system. In contrast, the nano-protrusions grown at step edges of $\sqrt{7}\times\sqrt{3}$ are non-superconducting, as discussed in the main manuscript. It can be seen directly from the fig.3d as well as from the large scale fig.5a and 5b that the characteristic scale of the inverse proximity lies always in the range 5-20 nm. Moreover as expected from inverse proximity effect, the larger the nanoprotrusion is, the more subgap states are filled in the vicinity of the step edge, thus suppressing superconductivity over a larger distance. The origin of such a strong difference in the superconducting properties of the nano-protrusions of both systems remains unclear at present. We anticipate however that this difference should be linked to the excess (in SIC) or to the deficiency (in $\sqrt{7}\times\sqrt{3}$) of Pb-atoms in these particular regions. In the $\sqrt{7}\times\sqrt{3}$ system the nano-protrusions regions would would be deficient of Pb atoms which are substituted by Si adatoms, because the latter become free after the removal of the initial $7 \times 7$ reconstruction and can easily diffuse to step edges. (see the structural part 1 above in the supplementary material). This process would result in a local reduction of the electronic density of states. Increased disorder combined to reduced metallicity would then explain the non-superconducting behaviour of the nano-protrusions in $\sqrt{7}\times\sqrt{3}$. In addition, this latter linear chain structure is poorly connected at step edges both because of the rougher step edges as a result of the nano-protrusions and the presence of domain rotation (see part one above and Fig.S3a). As a consequence, adjacent atomic terraces are only weakly linked in $\sqrt{7}\times\sqrt{3}$ whereas they are strongly coupled at step edges in SIC.
\vspace{\baselineskip}

4. \textit{Magnetic field response of the SIC system}. In parallel to the study carried out for the $\sqrt{7}\times\sqrt{3}$ Pb/Si(111) (see the main manuscript), we carefully examined the role of atomic step edges and stacking fault lines of the substrate in the magnetic field response of the SIC phase. Individual vortices were easily observed in direct scanning tunneling spectroscopy maps, as no peculiar region where superconductivity is weakened could be observed at zero-field. The vortex cores have a regular round shape as can be seen in Fig.4 of the main text). In contrast to the $\sqrt{7}\times\sqrt{3}$ phase (see Fig.5 of the main text), the vortices in the SIC sample were found to be located both in the middle of terraces and at step edges without a clear preference. Thus in the SIC, step edges and substrate stacking fault lines neither play the role of strong Josephson barriers nor that of vortex pinning centers. At higher fields, a vortex lattice is observed, as shown in Fig.4c. The lattice is disordered (as it is the case in many two-dimensional superconductors) but it is characterized by a regular inter-vortex spacing. The density of vortices in spectroscopic maps corresponds well to the magnetic flux crossing the sample.
\vspace{\baselineskip}

5. \textit{Theoretical part}. Here we derive the shape of the coherence peak in the density of states in a toy model of a small weakly disordered superconducting grain. We assume the following hierarchy of the length scales $L\gg l\gg\lambda_{F}$, where $L$ is the size of the grain, $l$ is the mean free path of the electrons and $\lambda_{F}$ is their Fermi wavelength. Under these conditions the Thouless conductance, $G$ (the sheet conductance in units of $e^{2}/2\pi\hbar$) is large, i.e. the Thouless energy $E_{T}=D/L^{2}$ is much larger than the mean energy spacing, $\delta=\overline{\epsilon_{a+1}-\epsilon_{a}}$ of the one-particle levels. Moreover we assume that $E_{T}\gg\Delta\gg\delta$ where $\Delta$ is superconducting gap.

Before we discuss the formation of the superconductivity in the grain, it will be useful to consider the toy model and discuss its relation to the grain afterwards. The toy model contains a set of $N_{D}\gg1$ one electron states, $\left|i\right\rangle $, $i=1\dots N_{D}$, chosen in such a way that the interaction between the electrons is local in this basis, e.g. $\left|i\right\rangle $ are on site states on a lattice and the interaction is short-range. A many-electron system is controlled by the Hamiltonian
\begin{equation}
H=H_{0}+H_{ee}=\sum_{ij}H_{ij}^{0}a_{i\sigma}^{\dagger}a_{j\sigma}-\lambda N_{D}\delta\sum_{i}a_{i\uparrow}^{\dagger}a_{i\downarrow}^{\dagger}a_{i\downarrow}a_{i\uparrow}\label{eq:H}
\end{equation}

Here the operator $a_{i\sigma}^{\dagger}$ ($a_{i\sigma}$) creates (annihilates) an electron with the spin $\sigma$ at the state $i$ and $\lambda$ is the dimensionless coupling constant. The orbitals, the eigenvectors $\left|a\right\rangle $ of the matrix $H_{ij}^{0}$ with the eigenvalues $\epsilon_{a}$ form a band; the width of this band $W=N_{D}\delta$ , plays the role of the Debye energy in the conventional theory of the superconductivity.

Matrix elements $\left\langle a,b\left|H_{ee}\right|c,d\right\rangle $ of the interacting part of the Hamiltonian between the pairs of the orbitals $\left|a,b\right\rangle =\left|a\right\rangle \left|\otimes b\right\rangle $ are given by
\[
\left\langle a,b\left|H_{ee}\right|c,d\right\rangle =-\lambda N_{D}\delta\sum_{i}\Psi_{a}^{*}(i)\Psi_{b}^{*}(i)\Psi_{c}(i)\Psi_{d}(i)
\]

It is convenient to split $H_{ee}$ into four parts:

\begin{subequations}

\begin{eqnarray}
H_{BCS}\! & =\! & -\lambda\delta\sum_{ab}\Lambda_{ab}a_{a\uparrow}^{\dagger}a_{a\downarrow}^{\dagger}a_{b\downarrow}a_{b\uparrow}\quad\Lambda_{ab}=N_{D}\sum_{i}\left[\Psi_{a}^{*}(i)\Psi_{b}(i)\right]^{2}\label{eq:H_BCS}\\
H^{(2)} & = & -\lambda\delta\sum_{ab}\left(K_{abbb}a_{a\uparrow}^{\dagger}a_{b\downarrow}^{\dagger}a_{b\downarrow}a_{b\uparrow}+K_{aaab}a_{a\uparrow}^{\dagger}a_{a\downarrow}^{\dagger}a_{a\downarrow}a_{b\uparrow}+\dots\right)\label{eq:H^(2)}\\
H^{(3)} & = & -\lambda\delta\sum_{abc}\left(K_{abcb}a_{a\uparrow}^{\dagger}a_{b\downarrow}^{\dagger}a_{c\downarrow}a_{b\uparrow}+\dots\right)\label{eq:H^(3)}\\
H^{(4)} & = & -\lambda\delta\sum_{abcd}K_{abcd}a_{a\uparrow}^{\dagger}a_{b\downarrow}^{\dagger}a_{c\downarrow}a_{d\uparrow}\label{eq:H^(4)}\\
K_{abcd} & = & N_{D}\sum_{i}\Psi_{a}^{*}(i)\Psi_{b}^{*}(i)\Psi_{c}(i)\Psi_{d}(i)\label{eq:K_abcd}
\end{eqnarray}

\end{subequations}

In the BCS approximation the full Hamiltonian $H_{ee}$ is replaced by $H_{BCS}$, (\ref{eq:H_BCS}), while the contributions (\ref{eq:H^(2)}-\ref{eq:H^(4)}) are neglected. In this approximation the gap equation

\begin{eqnarray}
\Delta_{a} & = & \lambda\delta\sum_{b}\Lambda_{ab}\frac{\tanh\beta E_{b}}{E_{b}}\label{eq:Delta_a}\\
E_{b} & = & \sqrt{\epsilon_{b}^{2}+\Delta_{b}^{2}}
\end{eqnarray}

becomes the conventional self-consistency equation, except that different orbitals, $a$, are strictly speaking characterized by different gaps, $\Delta_{a}$. However if the matrix $H_{ij}^{0}$ belongs to the Dyson orthogonal ensemble the components of its eigenvectors $\Psi_{a}(i)$ are real and not correlated with each other. The square of each component $\Psi_{i}^{2}(i)$ can be written as $\Psi_{i}^{2}(i)=N_{D}^{-1}+\delta_{a}^{i}$ where $\delta_{a}^{i}\sim N_{D}^{-1}$ has zero average $\sum_{i}\delta_{a}^{i}=0$ due to normalization condition. Because $\delta_{a}^{i}$ and $\delta_{b}^{j}$ are correlated only if $i=j$ and $a=b$, the sign of the product $\delta_{a}^{i}$$\delta_{b}^{i}$ randomly changes with $i$ and thus
\begin{equation}
\Lambda_{ab}=1+N_{D}\sum_{i}\delta_{a}^{i}\delta_{b}^{i}\approx1\label{eq:Lambda_ab}
\end{equation}

According to (\ref{eq:Lambda_ab}) in the leading order in $N_{D}^{-1}$ all $\Lambda_{ab}$ are the same for any pair $(a,b)$ and thus provided that $\Delta>\delta$ all orbitals have the same gap that can be determined from the usual self-consistency equation
\begin{equation}
\Delta=\lambda\delta\sum\frac{\tanh\beta E_{b}}{E_{b}}_{b}=\lambda\Delta\int_{0}^{W/2}d\epsilon\frac{\tanh\beta\sqrt{\epsilon^{2}+\Delta^{2}}}{\sqrt{\epsilon^{2}+\Delta^{2}}}\label{eq:GapEq}
\end{equation}

Why BCS approximation is good for weakly disordered superconductors? It turns out that the overlapping integrals $K_{abcd}$ (\ref{eq:K_abcd}) with at least one unpaired index are much smaller than $\Lambda_{ab}$ (\ref{eq:Lambda_ab}). For example, due to the orthogonality of the wave functions $\Psi_{a}(i)$ and $\Psi_{b}(i)$ $K_{aaab}$ can be presented as a sum over the sites $i$ of the terms that oscillate randomly from site to site:
\[
K_{aaab}=N_{D}\sum_{i}\Psi_{a}(i)\Psi_{b}(i)\delta_{a}^{i}
\]
so that $K_{aaab}\ll1$.

We now discuss the relation between the toy model in which all electron orbitals are described by Dyson ensemble and a small superconducting grain containing $N_{g}$ electrons. The random matrix decribes spectrum of the disordered grain only at the scale of Thouless energy $E_{T}=N_{T}\delta\ll W$. Moreover, for $\left|\epsilon_{a}-\epsilon_{b}\right|<E_{T}$ the wave functions $\Psi_{a}(i)$ ,$\Psi_{b}(i)$ are correlated on the number of sites $N_{g}/N_{T}=N_{g}\delta/E_{T}\gg1$. At larger $\left|\epsilon_{a}-\epsilon_{b}\right|>E_{T}$ the correlations decrease as negative power of $\left|\epsilon_{a}-\epsilon_{b}\right|$. In other words the correlations between wave functions at energies $\left|\epsilon_{a}-\epsilon_{b}\right|<E_{T}$ can be described by the effective model that contains only $N_{T}$ sites and orbitals with the bandwidth $E_{T}.$ In the following we focus on the states in this energy interval.

In the effective Hamiltonian restricted to the energy interval $\left|\epsilon_{a}-\epsilon_{b}\right|<E_{T}$ the effective interaction constant is renormalized by the neglected states outside of this energy interval. In the following we shall need energies $E_{T}\sim\Delta$ for which $\lambda_{eff}\sim1$.

The Hamiltonian $H^{(2)}$ leads to quasiparticle spead over many electron orbitals. Neglecting the processes that create other excitations, the quasiparticle is controlled by the effective Hamiltonian
\[
H_{qp}=\sum_{ab}b_{b\sigma}^{\dagger}b_{a\sigma}\left(\delta_{ab}-\lambda_{eff}\delta u_{a}u_{b}K_{ab}\right)
\]
where
\begin{align*}
K_{ab} & =\sum_{i}\left(\Psi_{a}(i)\Psi_{b}^{3}+\Psi_{b}(i)\Psi_{a}^{3}\right)\\
u_{a}^{2} & =\frac{1}{2}\left(1-\frac{\epsilon_{a}}{E_{a}}\right)
\end{align*}

In random matrix model all wave function are random with $\left\langle \Psi^{2}\right\rangle =1/N_{T},$ so the matrix elements are random variables with the zero average and
\[
\left\langle K_{ab}^{2}\right\rangle =48\lambda^{2}/N_{T}^{3}
\]

Non-diagonal matrix elements in the Hamiltonian (\ref{eq:H^(2)}) lead to the self energy correction to the quasiparticle Green function
\begin{equation}
\Sigma_{a}(\omega)=u_{a}^{2}\sum_{b}\frac{u_{b}^{2}K_{ab}^{2}}{\omega-E_{b}-\Sigma_{b}(\omega)}\label{eq:Sigma_Eq}
\end{equation}
The sum in (\ref{eq:Sigma_Eq}) runs over many states, so the result depends weakly on the index $a$, neglecting this dependence and introducing the dimensionless variable, $\Xi$ by $\Sigma_{a}(\epsilon)=u_{a}^{2}\Xi(\epsilon/\Delta)\Delta$ we get
\begin{eqnarray}
\Xi(x) & = & g\int d\xi\frac{u^{2}(\xi)}{x-\sqrt{1+\xi^{2}}-u^{2}(\xi)\Xi(x)}\label{eq:Xi_Eq}\\
u^{2}(\xi) & = & \frac{1}{2}\left(1-\frac{\xi}{\sqrt{1+\xi^{2}}}\right)\nonumber \\
g & = & \frac{48\lambda^{2}\delta^{2}}{\Delta^{2}}\nonumber
\end{eqnarray}

At small $g\ll1$ the self energy correction becomes irrelevant at $(\epsilon-\Delta)/\Delta\gg g^{2/3}.$ At these energies one gets conventional square root dependence for the density of states, $\nu(\epsilon)=\nu_{0}\sqrt{\Delta/(\epsilon-\Delta)}$. At smaller $(\epsilon-\Delta)/\Delta\lesssim g^{2/3}$ the square root singularity of the density of states is cutoff by the finite imaginary self energy. At small ($\epsilon-\Delta)/\Delta$ the behavior of the self energy is described by the simplified equation in which we neglect the BCS coherence factors dependence on energy
\begin{equation}
\Xi(x)=\frac{g}{2}\int\frac{d\xi}{x-\sqrt{1+\xi^{2}}-(1/2)\Xi(x)}=\frac{\pi ig}{\sqrt{2}\sqrt{x-1-(1/2)\Xi(x)}}\label{eq:Xi_eq_simplified}
\end{equation}
The Eq. (\ref{eq:Xi_eq_simplified}) shows that coherence peak saturates at $(\epsilon-\Delta)/\Delta\sim g^{2/3}$ due to $\Im\Xi\sim g^{2/3}$. Below the gap the imaginary part of the self energy rapidly decreases, the self energy becomes purely real at
\begin{equation}
\frac{\epsilon-\Delta}{\Delta}\leq-3\left(\frac{\pi g}{4\sqrt{2}}\right)^{2/3}\label{eq:Threshold}
\end{equation}
that describes the small shift of the gap edge. Density of states for different values of $g$ is shown in Fig.S4. One sees that even a small value of the dimensionless parameter $g$ results in a significant broadening of the BCS coherence peak.

The derivation and the result (\ref{eq:Xi_Eq}-\ref{eq:Threshold}) are appropriate for the small grain of superconductor. As explained above we expect that a bulk two dimensional material shows very similar behavior to the grain with $\delta=\Delta/G$. For this grain $g\sim50/G^{2}$.

In order to compare the expected density of states with the observations one needs to take into account a few effects. First, the actual experiments are performed at non-zero temperature which leads to a small but observable smearing of the gap edge. Similar smearing is caused by non-zero dephasing rate. Finally, the derivation above did not include the effect of the non-BCS terms (\ref{eq:H^(2)}-\ref{eq:H^(4)}) on the gap value. Qualitatively, these terms should increase pairing and thus enhance the gap at smaller $G$. Assuming that the value of the electron temperature is $T=0.1\Delta$ we find that the experimental data are best fit assuming very small dephasing rate $\tau_{\phi}^{-1}=0.05\Delta$ and values of $g=0.01$ and $g=0.03$ for the curves shown in the main text as shown in Fig.2 and S5. These values of $g$ correspond to the local conductance $G\sim30-100$ which is very reasonable for the weakly disordered film of atomic thickness.

\vspace{\baselineskip}

\vspace{\baselineskip}
\textbf{Acknowledgments}

The critical reading of our manuscript by Profs. G. Deutscher and N. Trivedi is greatefully acknowledged. This work was supported by University Pierre et Marie Curie UPMC "Emergence" project, French ANR Project 'ElectroVortex' and CNRS PICS funds, also partially funded by US-DOE grant DE-AC02-07CH11358.

\vspace{\baselineskip}
\textbf{Authors Contributions}

\vspace{\baselineskip}
\textbf{Additional information}

Supplementary information accompanies this paper.

\begin{figure*}
\includegraphics[width=\textwidth]{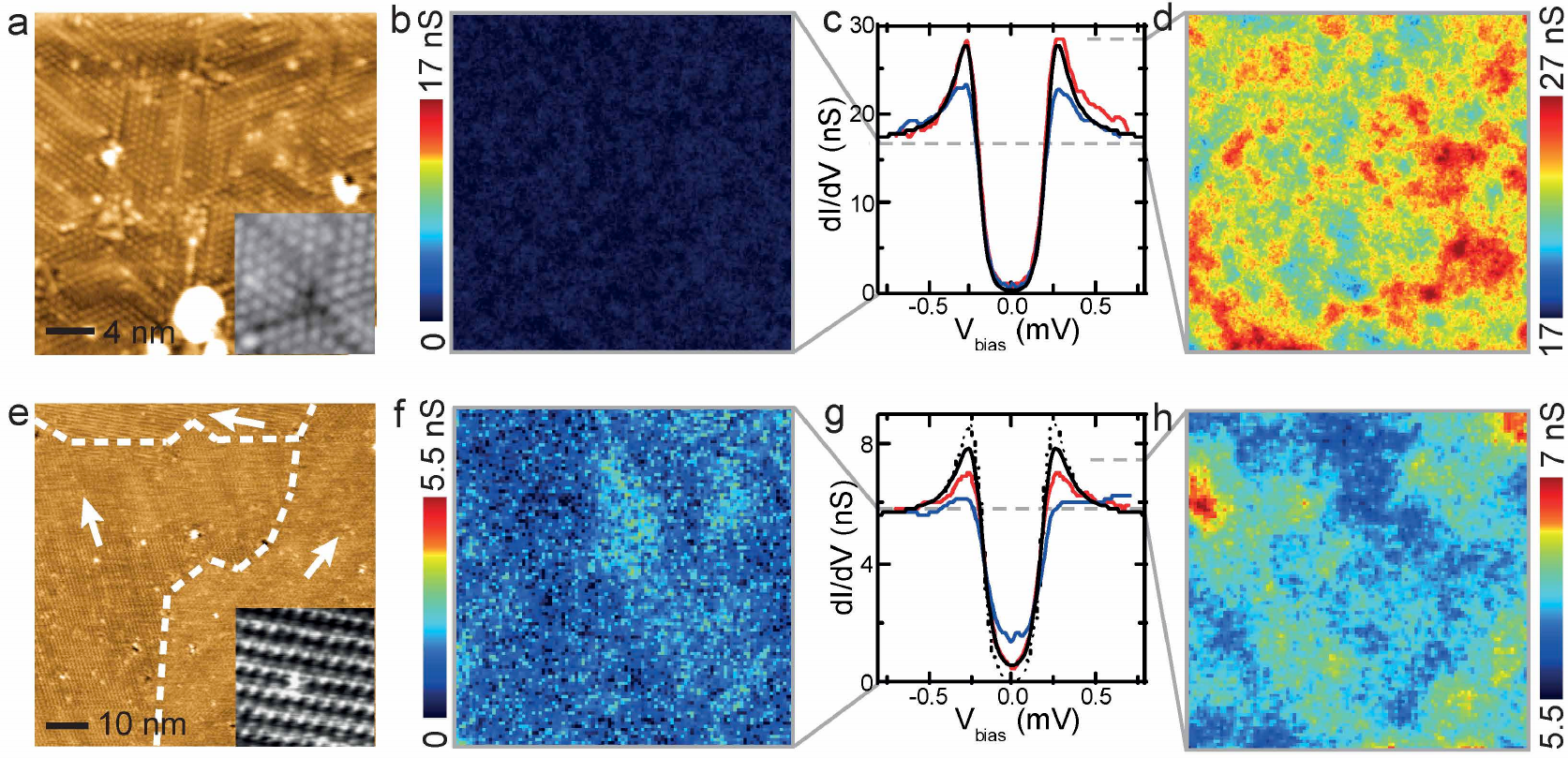}
\caption{\setlength{\baselineskip}{0.5\baselineskip} \textbf{Effect of disorder on the superconducting properties of one-atomic-layer films of Pb on Si(111),  at $T=320 $mK}. The properties of the SIC system are presented in the top pannel, those of the $\sqrt{7}\times\sqrt{3}$ one in the bottom pannel. The SIC system presents the characteristics of a 2D superconductor in the presence of a small local disorder, remaining reasonably BCS. The $\sqrt{7}\times\sqrt{3}$ shows strong departure from BCS behaviour. Its characteristics reveal a 2D superconductor fragilized by moderate disorder. (a),(e) - Scanning tunneling microscopy images (26nm$\times$26nm, 85nm$\times$85nm) acquired on atomic terraces. Adatoms, vacancies, individual domains and twin boundaries are visible; The dashed lines in (e) delimits three different rotational domains indicated by arrows. inserts show the atomic resolution. (b),(f) - The zero-bias conductance maps demonstrate that the superconducting energy gap exists everywhere on terraces, regardless of local and extended defects. However, while the zero-bias conductance is homogeneous in the SIC, it undergoes large spatial variations in $\sqrt{7}\times\sqrt{3}$ emphasizing subgap filling. (c),(g) - Representative local tunneling conductance spectra, showing that the amplitude of the quasiparticle peaks varies spatially. The quasiparticle peaks are heavily smeared out in the $\sqrt{7}\times\sqrt{3}$. The red (respectively blue) curve was acquired in a region with high (resp. low) coherence peaks, i.e. in yellow-red (blue-green) patches seen in (d),(h); the thin solid lines are best fits of the spectra using BCS formalism ($\Delta_{SIC}=0.23$meV, $\Delta_{\sqrt{7}\times\sqrt{3}}=0.20$meV). No broadening parameter is needed for the SIC, while in $\sqrt{7}\times\sqrt{3}$ the dotted line shows the pure BCS fit and the solid line is obtained with a large broadening parameter (d),(h) - Conductance maps at biases $V_{SIC}=0.29$mV and $V_{\sqrt{7}\times\sqrt{3}}=0.26$mV, corresponding to the energy of the quasiparticle peaks in the tunneling spectra (c),(g). The spatial variations of the peaks amplitude do not follow the local structural defects observed in the topography; rather, they evolve on a scale of few nm, which is much smaller than the coherence length of both systems close to 50nm.} \label{Fig1}
\end{figure*}

\begin{figure*}
\includegraphics[width=\textwidth]{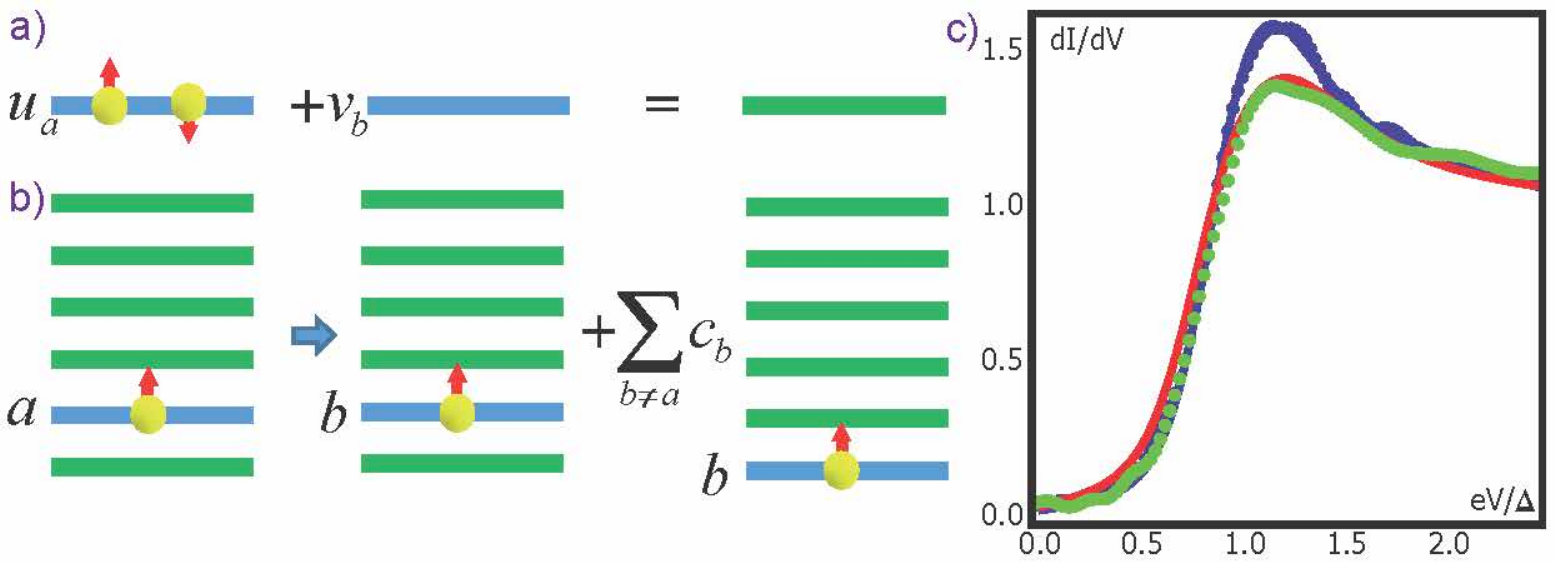}
\caption{\small \textbf{Cartoon of the BCS theory and leading corrections.} (a) Upper diagram: schematics of the states in conventional BCS theory, in the leading approximations all electron orbitals are independent. (b) Lower diagram: a moderate disorder admixes the blocked states containing one unpaired electron as shown in lower diagram resulting in the smearing of the coherence peak. (c) Fitting of the SIC tunneling spectra.} \label{Fig2}
\end{figure*}

\begin{figure}
\includegraphics[width=10.5cm]{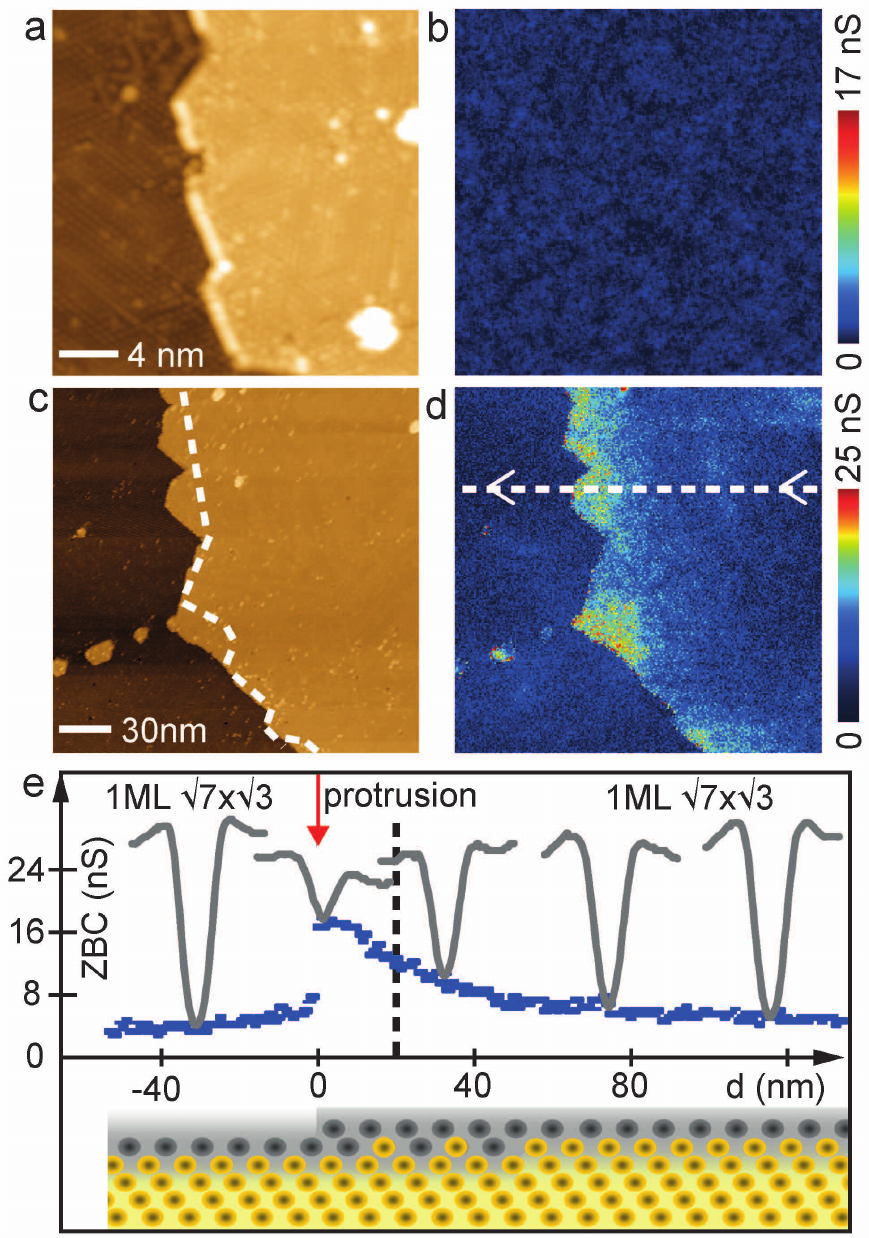}
\caption{\setlength{\baselineskip}{0.5\baselineskip} \textbf{Superconducting properties of one-atomic-layer films of Pb on Si(111) near step edges, at $T=320 $mK.} While the superconducting order in the SIC remains insensitive to monoatomic step edges, the order parameter is disrupted in $\sqrt{7}\times\sqrt{3}$ on a subnanometer length right at the step edge. Combined with magnetic field measurement (see figures 3 and 4) this shows that a monoatomic step is a Josephson junction in $\sqrt{7}\times\sqrt{3}$. (a),(c) - Respective topographic images (26nm$\times$26nm) of SIC and $\sqrt{7}\times\sqrt{3}$ (210nm$\times$210nm) centered at single atomic step edges. The dashed line in (c) delimits the separation between the nano-protrusions and the upper terrace to which they are attached; (b),(d) - Zero-bias conductance maps corresponding respectively to regions shown in (a) and (c). In (b) for the SIC sample, the local tunneling spectra exhibit a well developed superconducting gap across the step edge which is indistinguishable in the spectroscopic map. In (d), the nano-protrusions in $\sqrt{7}\times\sqrt{3}$ exhibit a high zero-bias conductance and thus are not superconducting. In the vicinity of the nano-protrusions, the upper terrace shows a spatially dependent zero-bias conductance due to inverse proximity effect. The superconducting gap is restored on a scale of several tens of nm. On the lower terrace, the gap abruptly reopens in the immediate vicinity of the step edge close to the nano-protrusion. (e) Spatial evolution of the zero-bias conductance along the dashed line shown in (d), when going from the upper to the lower terrace through a nano-protrusion. The lateral extension of the nano-protrusion is between the red arrow (0 nm) and the black dashed line (20 nm) as seen in Fig.2c. Selected $dI/dV$  spectra measured along this dashed line are shown in real conductance units. Yellow and grey disks schematically depict silicon and lead atoms in the vicinity of the step edge.} \label{Fig2}
\end{figure}

\begin{figure*}
\includegraphics[width=\textwidth]{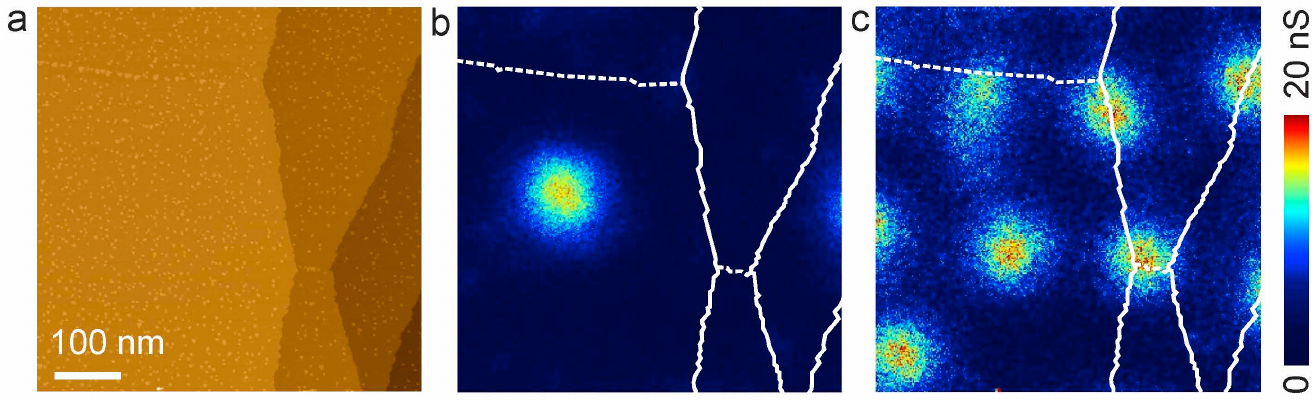}
\caption{\small \textbf{Magnetic field response of the striped incommensurate Pb/Si(111) phase.} (a) Topographic image (600nm$\times$600 nm) showing single atomic steps, decorated stacking fault lines of the substrate and very small 1ML high nanoislands. (b,c) Corresponding zero-bias conductance maps acquired at $T=320 $mK under magnetic field: (b) - 0.01T, (c) - 0.04T. The monoatomic step edges visible in the topography are indicated by continuous lines and the decorated stacking fault lines of the substrate are shown by dashed lines. Vortex cores are visualized as regions where the superconducting gap is suppressed and appear in yellow. Abrikosov-type of vortices are directly seen in these zero-bias conductance maps.} \label{Fig3}
\end{figure*}

\begin{figure}
\includegraphics[width=\columnwidth]{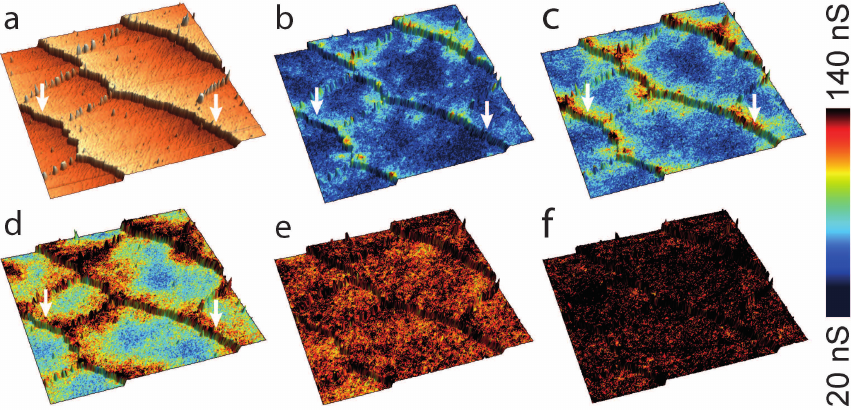}
\caption{\small \textbf{Revealing vortices in $\sqrt{7}\times\sqrt{3}$ Pb/Si(111).} Top panel: Zero-bias tunneling conductance maps of a 600nm$\times$600nm region of $\sqrt{7}\times\sqrt{3}$ Pb/Si(111), at $T=320 $mK under a magnetic field of (a) - $B_1$=0T; (b) - $B_2=0.04$T; (c) - $B_3=0.08$T; bottom panel: difference maps showing the locations at which superconductivity has been suppressed in (d) between $B_1$ and $B_2$, in (e) between $B_2$ and $B_3$. (d-e) - This analysis allows us to visualize vortex cores appearing in red, occupying first the weak links regions constituted by step edges. In d and e, the step edges (substrate stacking fault lines) are indicated by continuous (dashed) lines.} \label{Fig4}
\end{figure}

\begin{figure*}
\includegraphics[width=\textwidth]{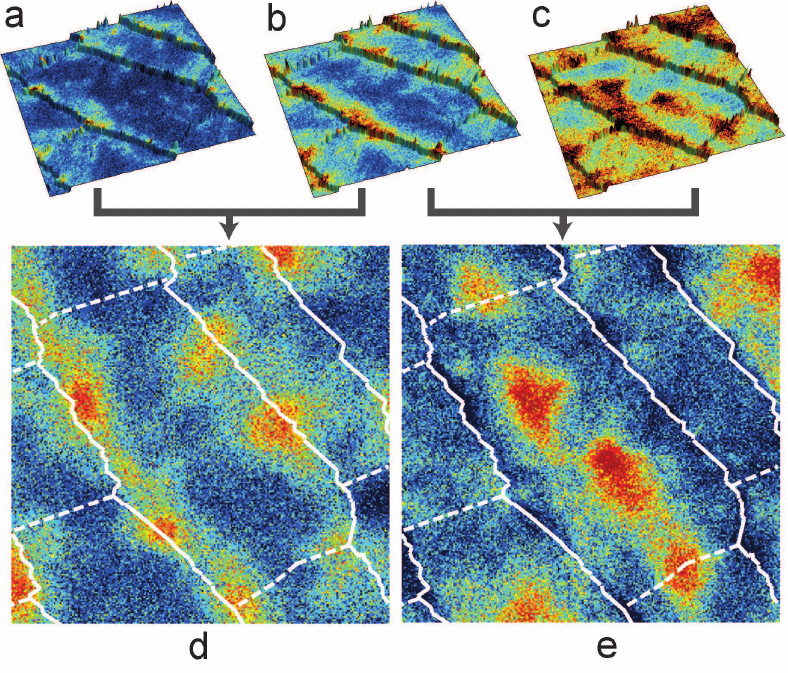}
\caption{\small \textbf{Magnetic field response of $\sqrt{7}\times\sqrt{3}$ Pb/Si(111): revealing Josephson junctions.} (a) - topographic image (600nm$\times$600nm) showing single atomic steps and decorated stacking fault lines of the substrate; (b-f) - corresponding zero-bias conductance maps acquired at 0T, 0.04T, 0.08T, 0.16T and 0.24T respectively. (b) - at zero field most of the sample is superconducting except at nano-protusions and at decorated stacking fault lines of the substrate; (c-d) - when increasing the field, superconductivity weakens preferentially in the vicinity of the nano-protusions, the decorated stacking fault lines, and also remarkably at step edges. (e) - superconductivity survives only in very small patches, located on terraces, and attached to ascending step edges (f) - the sample recovers the normal state. The imaging conditions are ($V=-50$ mV, $I=0.05$ nA).} \label{Fig5}
\end{figure*}

\begin{figure*}
\includegraphics[width=\textwidth]{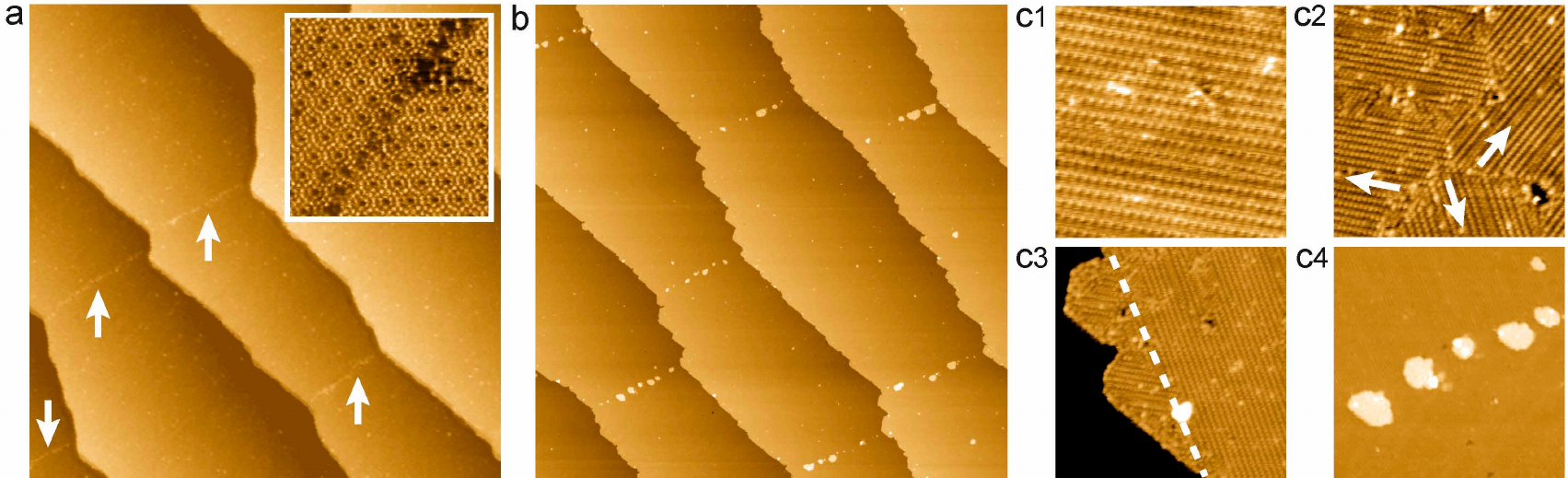}
\caption{\small \textbf{Supplementary Figure S1. Structural properties of the $\sqrt{7}\times\sqrt{3}$ Pb/Si(111) sample}. (a) - topographic constant current scanning tunneling microscopy  image (800nm$\times$800nm) of the 7$\times$7 Si(111) substrate. It reveals atomically clean large terraces separated by single atomic steps; white arrows point on the stacking fault lines of the 7$\times$7 reconstruction crossing individual terraces; inset: zoom on a stacking fault line of the $7\times7$ reconstruction; (b) - topographic image (800nm$\times$800nm) of the $\sqrt{7}\times\sqrt{3}$ Pb/Si(111) sample. Atomically clean terraces are preserved. The step edges appear more irregular; (c1 - c4) - topographic images - zooms on different regions of (b). (c1) - atomic resolution of the $\sqrt{7}\times\sqrt{3}$ Pb/Si(111) reconstruction; (c2) - boundaries separating the three orientational domains existing on the reconstructed surface. Each orientation is indicated by an arrow; (c3) - nano-protrusions decorating the step edges (the step edge is indicated by a dashed line); (c4) - nano-protrusions nucleated at the stacking fault lines of the substrate. The imaging conditions are ($V=+1.5$ V, $I=0.1$ nA) for image (a) and ($V=-50$ mV, $I=0.05$ nA) for image (b).} \label{FigS1}
\end{figure*}

\begin{figure*}
\includegraphics[width=\textwidth]{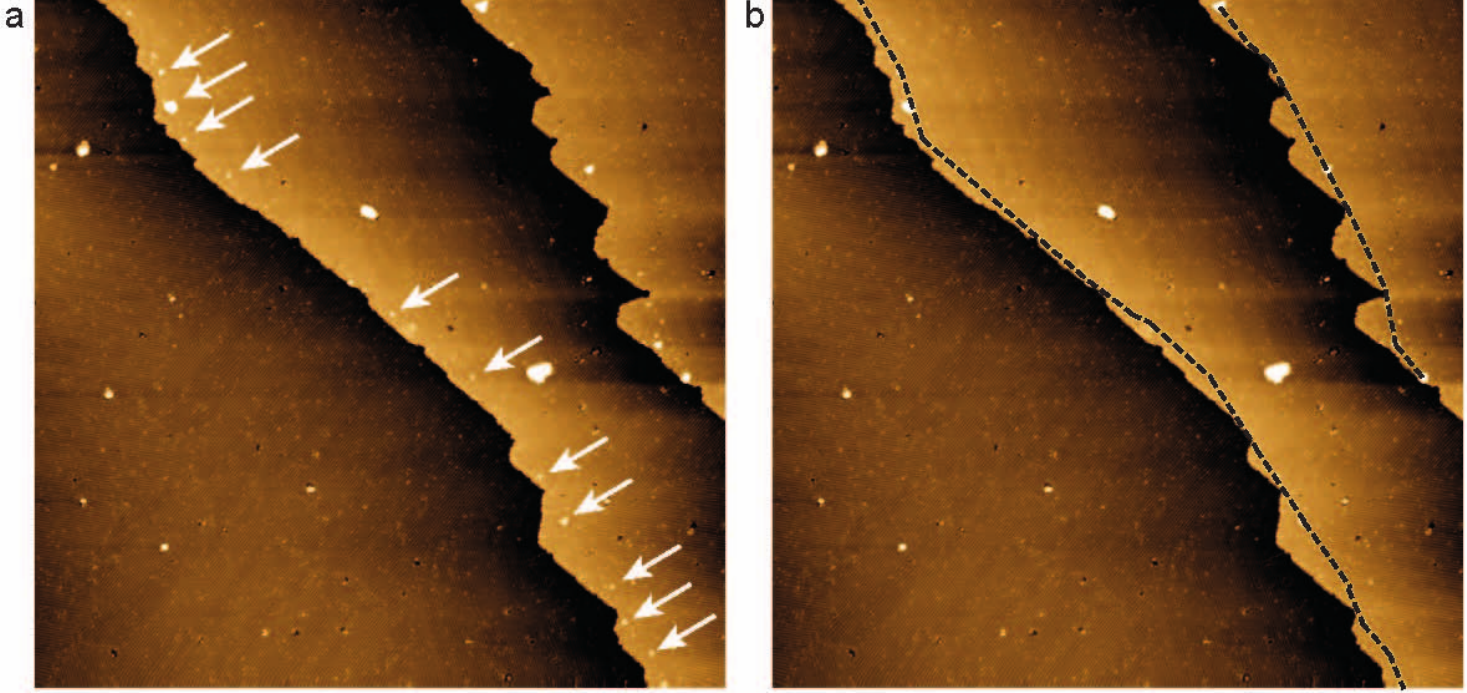}
\caption{\small \textbf{Supplementary Figure S2. Determining the Si step edge near the overhanging nanoprotrusion in $\sqrt{7}\times\sqrt{3}$ Pb/Si(111)}. (a) - topographic constant current scanning tunneling microscopy  image (210nm$\times$210nm) of the $\sqrt{7}\times\sqrt{3}$ Pb/Si(111) sample. Three atomic terraces are revealed. On the two upper terraces, small clusters/nanoislands are seen as white dots (some of these are indicated by arrows) and decorate the steps all along their edges. Through these lying nanostructures we draw a path, indicated by a dashed line in b), which delimitates the Si step edge location. During the growth of the $\sqrt{7}\times\sqrt{3}$ Pb/Si(111), the material which has been removed from the Si(111)-$7\times7$ (the top Si adatoms forming the $7\times7$ reconstruction) has diffused toward the Si step edge and is mixed with Pb adatoms. Thus the regions located between the dashed lines and the surface step edge constitute what we call the nanoprotrusions. We believe that these latter regions are deficient of Pb adatoms with respect to the standard Pb coverage of the $\sqrt{7}\times\sqrt{3}$ phase, due to the intermixing with Si adatoms. The imaging conditions are (V=-0.13 V, I=0.1 nA).} \label{FigS2}
\end{figure*}

\begin{figure*}
\includegraphics[width=\textwidth]{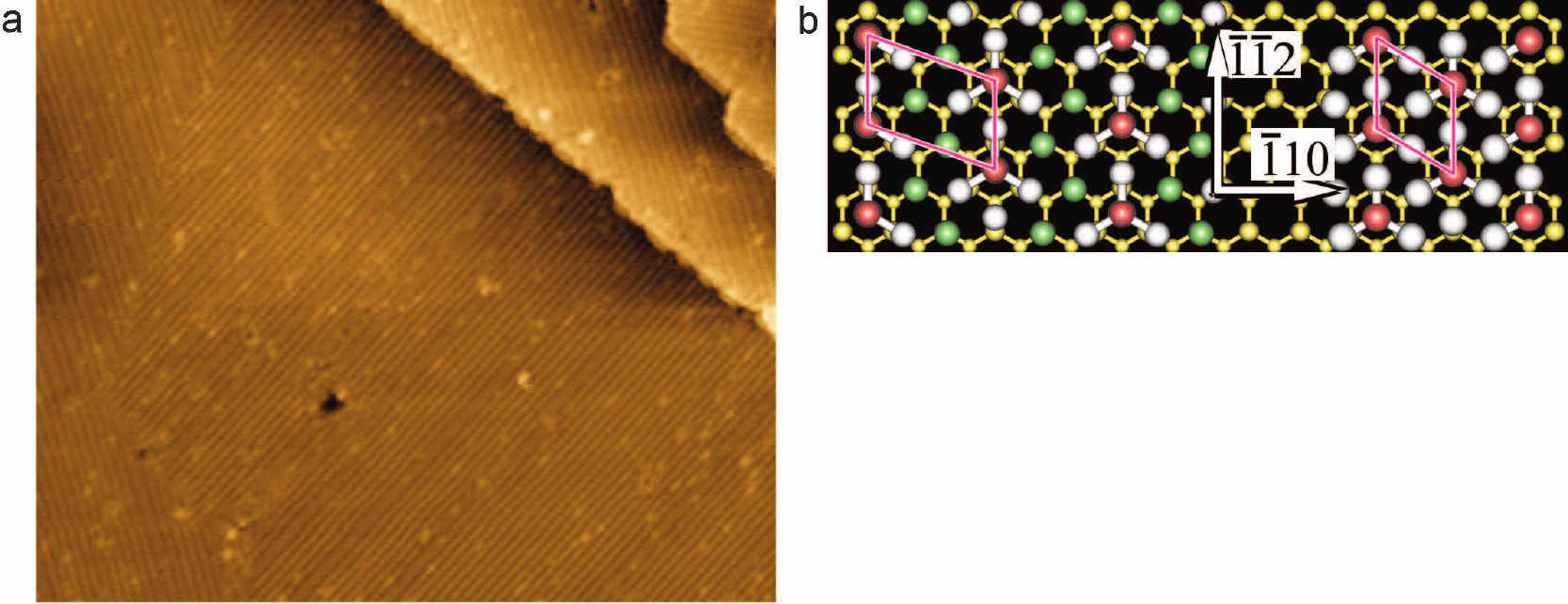}
\caption{\small \textbf{Supplementary Figure S3} (a) - topographic constant current scanning tunneling microscopy image (69nm$\times$69nm) of the $\sqrt{7}\times\sqrt{3}$ Pb/Si(111) sample showing the change in the orientation of the three equivalent domains on adjacent terraces. Most often for the $\sqrt{7}\times\sqrt{3}$ Pb/Si(111) system this situation is encountered, where at the step edge the $\sqrt{7}\times\sqrt{3}$ domains change orientation by 120 degrees across adjacent terraces. Togoether with the existence of overhanging nanoprotrusions at some locations along the step edges, this structural effect implies a poorer connection of the electronic properties of adjacent terraces at the edge, the edge becoming a Josephson junction in the superconducting state (see Fig. 4 and 5 in the main text). The imaging conditions are (V=-0.13 V, I=0.1 nA). (b) schematics of the atomic structures of the $\sqrt{7}\times\sqrt{3}$ Pb/Si(111) (left) and dense $\alpha-\sqrt{3}\times\sqrt{3}$ Pb/Si(111) (right). The SIC phase is obtained by combining unit cells having the dense $\alpha-\sqrt{3}\times\sqrt{3}$ periodicity with meandering walls made of $\sqrt{7}\times\sqrt{3}$ unit cells. Yellow atoms represent Si atoms, Red: Pb atoms on hollow sites H3 (no Si atom below), Green: Pb atoms on top sites T4 (top Si atom below), White: the non-symmetry Pb atoms. The respective unit cells are indicated by red continuous lines.} \label{FigS3}
\end{figure*}

\begin{figure}
\includegraphics[width=0.8\columnwidth]{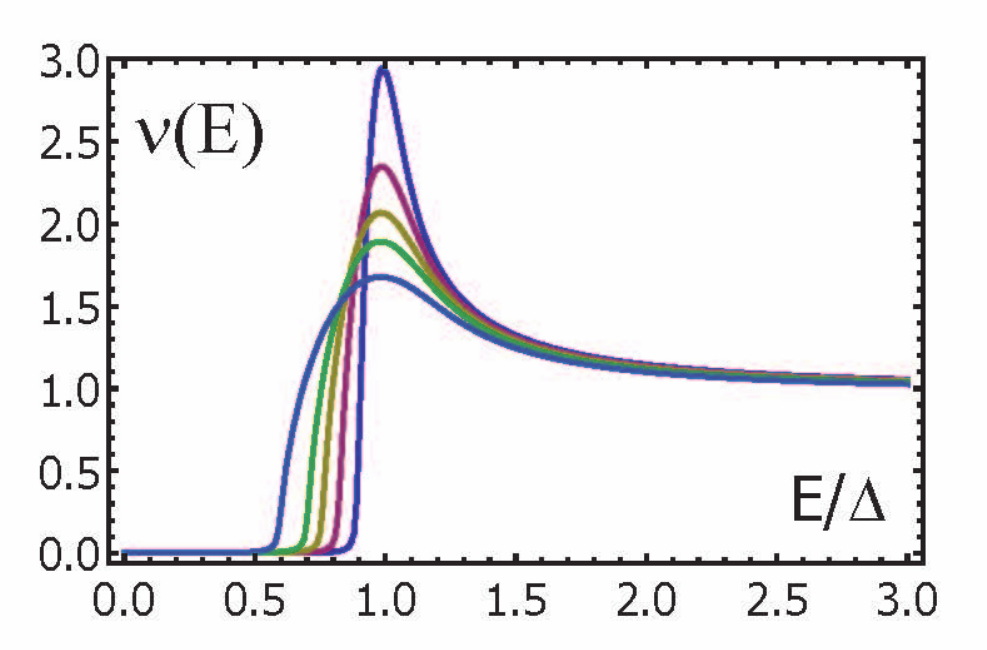}
\caption{\small \textbf{Supplementary Figure S4.} Average density of states obtained by solving numerically (\ref{eq:Xi_Eq}) and inserting it in $\nu(\epsilon)=\Im G(\epsilon)$ for different values of $g=0.01-0.06$. }
\label{fig:S_DOS}
\end{figure}

\begin{figure}
\includegraphics[width=0.8\columnwidth]{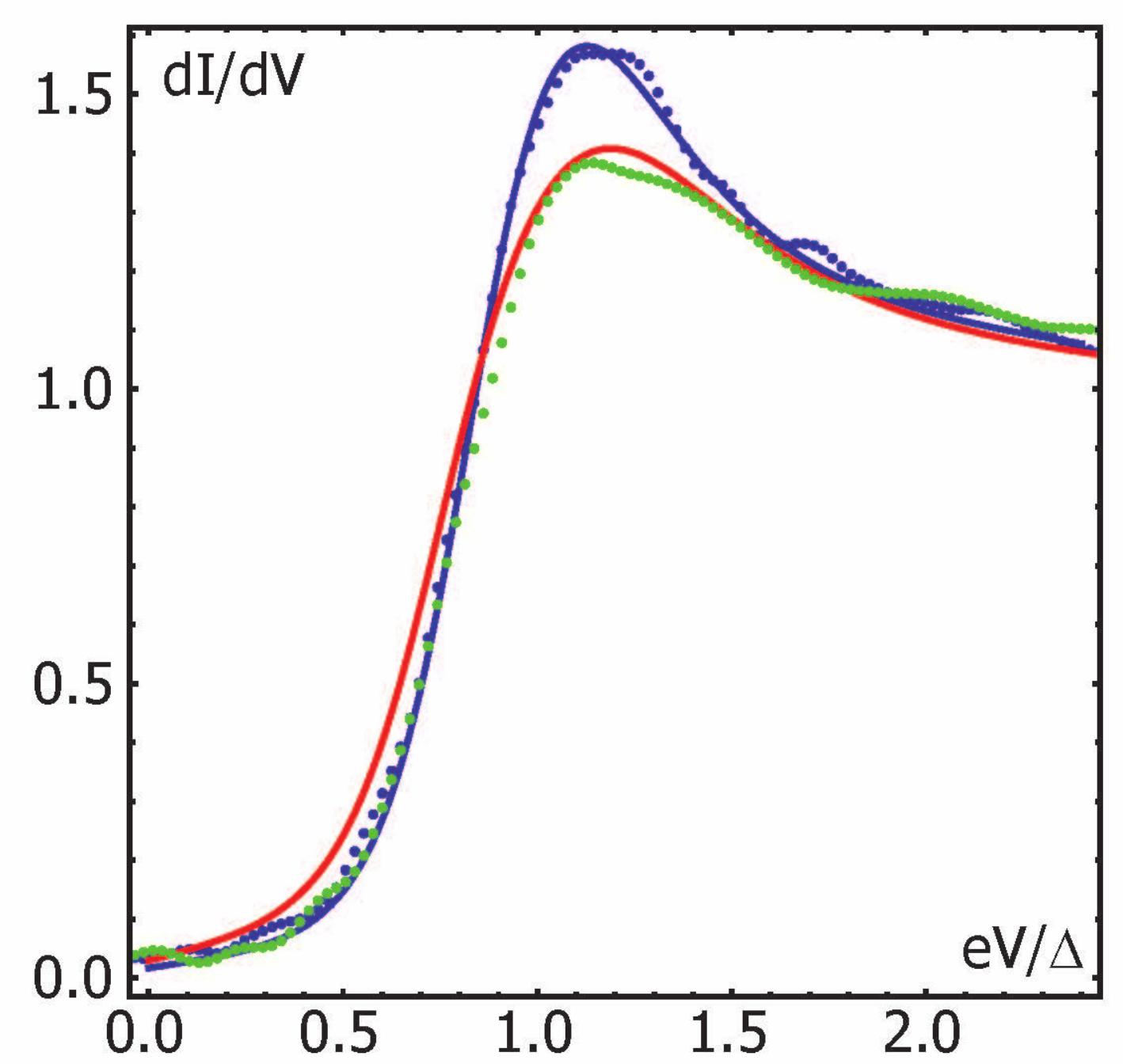}
\caption{\small \textbf{Supplementary Figure S5.} Fit of the experimental data to the expected $dI/dV$ that takes into account non-zero temperature, small dephasing rate and gap variations. The blue and red curves are the theoretical expectations for different values of $g=0.03$ and $g=0.05$ respectively. The best fit is obtained if one assumes that the bare gap value is slightly larger $\Delta_{g=0.03}=1.05\Delta{}_{g=0.01}$ for the more disorder location. In all curves $T=0.1\Delta$ and $\tau_{\phi}^{-1}=0.05\Delta$. The quality of the fit is only weakly sensitive to the actual values of $T$ and $\tau_{\phi}^{-1}$: one can produce very similar fits assuming slightly higher temperature and lower dephasing $\tau_{\phi}^{-1}$. }
\label{fig:DataFit}
\end{figure}


\textbf{References}

\begin{thebibliography}{00}

\bibitem{Anderson} Anderson, P. W. Theory of dirty superconductors. \textit{J. Phys. Chem. Solids} \textbf{11,} 26 (1959).

\bibitem{Goldman1998} Goldman, A. M. \& Markovic, N. Superconductor-insulator transitions in the two-dimensional limit. \textit{Physics Today.} \textbf{51,} 11, 39 (1998).

\bibitem{Huscroft}  Huscroft, C. \& Scalettar, R. T. Evolution of the density of states gap in a disordered superconductor. \textit{Phys. Rev. Lett.} \textbf{81}, 2775 (1998).

\bibitem{Bouadim2011} Bouadim, K., Loh, Y. L., Randeria, M. \& Trivedi, N. Single- and two-particle energy gaps across the disorder-driven superconductor-insulator transition. \textit{Nature Phys.} \textbf{7,} 884 (2011).

\bibitem{Feigelman} Feigel'man, M. V. \& Skvortsov, M. A. Universal broadening of the Bardeen-Cooper-Schrieffer coherence peak of disordered superconducting films. \textit{Phys. Rev. Lett.} \textbf{109}, 147002 (2012).

\bibitem{Feigelman2} Feigel'man, M. V. Ioffe, L.B. Kravtsov, V.E. \& Cuevas, E. Fractal superconductivity near localization threshold. \textit{Annals of Physics} \textbf{325} 1390-1478 (2010).

\bibitem{Sacepe2011} Sac\'{e}p\'{e}, B. \textit{et. al.} Localization of preformed Cooper pairs in disordered superconductors. \textit{Nature Phys.} \textbf{7,} 239 (2011).

\bibitem{Xue} Zhang, T. \textit{et al.} Superconductivity in one-atomic-layer metal films grown on Si(111). \textit{Nature Phys.} \textbf{6,} 104 (2010).

\bibitem{Yamada} Yamada, Y., Hirahara, T., \& Hasegawa, S. Magnetoresistance measurements of a superconducting surface state of In-induced and Pb-induced structures on Si(111). \textit{Phys. Rev. Lett.} \textbf{110,} 237001 (2013).

\bibitem{Uchihashi} Uchihashi, T., Puneet, M., Aono, M.,\& Nakayama, T. Macroscopic superconducting current through a silicon surface reconstruction with indium adatoms: Si(111)-($\sqrt{7}\times\sqrt{3}$)-In. \textit{Phys. Rev. Lett.} \textbf{107,} 207001 (2011).

\bibitem{FeSeXue} Song, C. L. \textit{et al.} Suppression of superconductivity by twin boundaries in FeSe. \textit{Phys. Rev. Lett.} \textbf{109,} 137004 (2012).

\bibitem{Yang} Guo, Y. \textit{et al.} Superconductivity modulated by quantum size effects. \textit{Science} \textbf{306,} 1915 (2004).

\bibitem{Ozer} \"{O}zer, M. M., Thompson, J. R. \& Weitering, H. H. Hard superconductivity of a soft metal in the quantum regime. \textit{Nature Phys.} \textbf{2,} 173 (2006).

\bibitem{Eom} Eom, D. Qin, S. Chou, M.-Y. \& Shih, C. K. Persistent superconductivity in ultrathin Pb films: a scanning tunneling spectroscopy study. \textit{Phys. Rev. Lett.} \textbf{96,} 027005 (2006).

\bibitem{OzerAlloy} \"{O}zer, M. M., Jia, Y., Zhang, Z., Thompson J. R. \& Weitering, H. H. Tuning the quantum stability and superconductivity of ultrathin metal alloys. \textit{Science} \textbf{316}, 1594 (2007).

\bibitem{Brun} Brun, C. \textit{et al.} Reduction of the superconducting gap of ultrathin Pb islands grown on Si(111). \textit{Phys. Rev. Lett.} \textbf{102,} 207002 (2009).

\bibitem{Qin} Qin, S. Y., Kim, J., Niu, Q. \& Shih, C. K. Superconductivity at the two-dimensional limit. \textit{Science} 324, 1314-1317 (2009).

\bibitem{Devilstaircase} Hupalo, M., Schmalian, J. \& Tringides, M. C. Devil staircase in Pb/Si(111) ordered phases. \textit{Phys. Rev. Lett.} \textbf{90,} 216106 (2003).

\bibitem{Seehofer} Seehofer, L. Falkenberg, G. Daboul, D. \& Johnson, R. L. Structural study of the close-packed two-dimensional phases of Pb on Ge(111) and Si(111). \textit{Phys. Rev. B}, \textbf{51,} 13503 (1995).

\bibitem{Horikoshi}  Horikoshi, K. Tong, X. Nagao, T. \& Hasegawa S. Structural phase transitions of Pb-adsorbed Si(111) surfaces at low temperatures. \textit{Phys. Rev. B} \textbf{60,} 13287 (1999).

\bibitem{Kumpf}  Kumpf, C. \textit{et al.}, Structural study of the commensurate incommensurate low-temperature phase transition of Pb on Si(111). \textit{Surf. Sci.} \textbf{448}, L213-L219 (2000).

\bibitem{Brochard} Brochard, S. \textit{et al.}, Ab initio calculations and scanning tunneling microscopy experiments of the Si(111)$\sqrt{7}\times\sqrt{3}$-Pb surface. \textit{Phys. Rev. B}, \textbf{66,} 205403 (2002).

\bibitem{Cudazzo}  Cudazzo, P., Profeta, G. \&  Continenza, A. Low temperature phases of Pb/Si(111) and related surfaces. \textit{Surf. Sci.}, \textbf{602,} 747 (2008).

\bibitem{ARPES0} Choi, W. H., Koh, H., Rotenberg, E. \& Yeom, H. W. Electronic structure of dense Pb overlayers on Si(111) investigated using angle-resolved photoemission. \textit{Phys. Rev. B} \textbf{75,} 075329 (2007).

\bibitem{ARPES1} Kim, K. S., Jung, S. C., Kang, M. H. \& Yeom, H. W. Nearly massless electrons in the silicon interface with a metal film. \textit{Phys. Rev. Lett.} \textbf{104,} 246803 (2010).

\bibitem{KangDFT} Jung, S. C. \& Kang, M. H. Triple-domain effects on the electronic structure of Pb/Si(111)-($\sqrt{7}\times\sqrt{3}$): Density-functional calculations. \textit{Surf. Sci.} \textbf{605,} 551 (2011).


\bibitem{DeGennes} de Gennes, P. G. Ed. Superconductivity Of Metals And Alloys. (W. A. Benjamin Inc., New York, 1966).

\bibitem{Altshuler1982} Altshuler, B. L., Aronov, A.G. \& Khmelnitskii, D.E. Effects of Electron-electron Collisions with Small Energy Transfers on Quantum Localization. J.Phys. \textbf{C 15}, 7367 (1982).

\bibitem{Altshuler1985} Altshuler, B. L. \& Aronov, A.G. in Electron-Electron Interactions in Disordered Systems, ed. by A.L. Efros, M. Pollak (North Holland, 1985), 1.

\bibitem{Aleiner2000} Kurland, I. L., Aleiner, I. L. \& Altshuler, B. L. Mesoscopic magnetization fluctuations for metallic grains close to the Stoner instability. Phys Rev \textbf{B 62}, 14886-14897 (2000).

\bibitem{Thouless1974} Thouless, D. J. Electrons in disordered systems and the theory of localization. Physics Reports \textbf{13,} 93-142 (1974).

\bibitem{DilRashbaQWS} Dil, J. H. \textit{et al.} Rashba-Type Spin-Orbit Splitting of QuantumWell States in Ultrathin Pb Films, \textit{Phys. Rev. Lett.} \textbf{101,} 266802 (2008).

\bibitem{SetphanePons} Gierz, I., \textit{et al.} Silicon Surface with Giant Spin Splitting. \textit{Phys. Rev. Lett.} \textbf{103,} 046803 (2009).

\bibitem{RashbaMLPb} Yaji, K., \textit{et al.} T. Large Rashba spin splitting of a metallic surface-state band on a semiconductor surface. Nat. Commun. 1:17 doi: 10.1038 / ncomms1016 (2010).

\bibitem{DopRashbaQWS} Slomski, B., Landolt, G., Bihlmayer, G., Osterwalder, J. \& Dil, J. H. Tuning of the Rashba effect in Pb quantum well states via a variable Schottky barrier. Sci. Rep. 3, 1963; DOI:10.1038/srep01963 (2013).


\bibitem{Gor'kovRashba} Gor\'{}kov, L. P., and Rashba, E. I. Superconducting 2D System with Lifted Spin Degeneracy: Mixed Singlet-Triplet State. \textit{Phys. Rev. Lett.} \textbf{87,} 037004 (2001).



\bibitem{Josephson} Josephson, B. D. Possible new effects in superconductive tunnelling. \textit{Phys. Lett.} \textbf{1,} 251 (1962).


\bibitem{McMillan} McMillan, W. L. Tunneling model of the superconducting proximity effect. \textit{Phys. Rev.} \textbf{175,} 537 (1968).

\bibitem{SerrierGarcia} Serrier-Garcia, L. \textit{et al.} Scanning tunneling spectroscopy study of the proximity effect in a disordered tow-dimensional metal. \textit{Phys. Rev. Lett.} \textbf{110}, 157003 (2013).

\bibitem{Kim} Kim, J. \textit{et al.}, Visualization of geometric influences on proximity effects in heterogeneous superconductor thin films. \textit{Nature Phys.} \textbf{8,} 464 (2012).

\bibitem{Abrikosov} Abrikosov, A. A. On the magnetic properties of superconductors of the second group. \textit{Zh. Eksp. i Teor. Fiz.} 32, 1442 (1957). \textit{Soviet Physics JETP} \textbf{5,} 1174 (1957); Nobel lecture: type-II superconductors and the vortex lattice. \textit{Rev. Mod. Phys.} \textbf{76,} 975 (2004).

\bibitem{Hess} Hess, H. F., Robinson, R. B., Dynes, R. C., Valles, J. M. \& Waszczak, J. V. Scanning-tunneling-microscope observation of the Abrikosov flux lattice and the density of states near and inside a fluxoid. \textit{Phys. Rev. Lett.} \textbf{62,} 214 (1989).

\bibitem{Gurevich} Gurevich, A. Nonlinear viscous motion of vortices in Josephson contacts. \textit{Phys. Rev. B}, \textbf{48,} 12857 (1993).

\bibitem{Cren} Cren, T. Serrier-Garcia, L. Debontridder, F. \& Roditchev, D. Vortex fusion and giant vortex states in confined superconducting condensates. \textit{Phys. Rev. Lett.} \textbf{107,} 097202 (2011).

\bibitem{Hupalo2002} Hupalo, M., Chan, T. L., Wang, C. Z., Ho, K. M. \& Tringides, M.C. Atomic models, domain-wall arrangement, and electronic structure of the dense Pb/Si(111)-$\sqrt{3}\times\sqrt{3}$ phase. \textit{Phys. Rev. B} \textbf{66}, 161410(R), (2002).
\end{thebibliography}
\end{document}